\documentclass{agujournal2019}

\usepackage{url}
\usepackage{soul}
\usepackage{natbib}
\usepackage[none]{hyphenat}
\usepackage{amsmath}
\usepackage{booktabs}
\usepackage{multirow}
\usepackage{placeins}
\usepackage{subcaption}

\draftfalse

\journalname{JGR: Machine Learning and Computation}

\begin{document}

\title{CLARE: Classification-based Regression for Electron Temperature Prediction}

\authors{Michael Liang\affil{1}, Blake DeHaas\affil{1}, Naomi Maruyama\affil{1}, Xiangning Chu\affil{1}, Takumi Abe\affil{2}, Koh-Ichiro Oyama\affil{2}}

\affiliation{1}{Laboratory for Atmospheric and Space Physics, University of Colorado Boulder, Boulder, CO, USA}
\affiliation{2}{Japan Aerospace Exploration Agency (JAXA), Sagamihara, Kanagawa, Japan}

\correspondingauthor{Michael Liang}{michaelliang15@gmail.com}
\correspondingauthor{Naomi Maruyama}{naomi.maruyama@lasp.colorado.edu}

\begin{keypoints}
\item First ML model for plasmasphere electron temperature, achieving 69.67\% accuracy within 10\% tolerance versus 13.49\% for best baseline.
\item Classification-based regression improves accuracy 6.46\% over continuous regression while providing built-in uncertainty quantification.
\item CLARE captures solar storm dynamics at 46.17\% accuracy despite storms comprising only 0.74\% of training data.
\end{keypoints}

\begin{abstract}

Electron temperature ($T_e$) is an important parameter governing space weather in the upper atmosphere, but has historically been underexplored in the space weather machine learning literature. We present \textbf{CLARE}, a machine learning model for predicting electron temperature in the Earth's plasmasphere trained on AKEBONO (EXOS-D) satellite measurements as well as solar and geomagnetic indices. CLARE uses a classification-based regression architecture that transforms the continuous $T_e$ output space into 150 discrete classification intervals. Training the model on a classification task improves prediction accuracy by 6.46\% relative compared to a traditional regression model while also outputting uncertainty estimation information on its predictions. On a held out test set from the AKEBONO data, the model's $T_e$ predictions achieve 69.67\% accuracy within 10\% of the ground truth and 46.17\% on a known geomagnetic storm period from January 30th to February 7th, 1991. We show that machine learning can be used to produce high-accuracy $T_e$ models on publicly available data.

\end{abstract}

\section*{Plain Language Summary}
\label{plain_language_summary}
The Earth's plasmasphere is a region of cold plasma surrounding the Earth above the upper atmosphere. Electron temperature ($T_e$) is one characteristic of this plasma which critically influences space weather in the region. Accurate prediction of $T_e$ is vital for understanding near earth space environments where many satellites orbit, but remains challenging due to a lack of data and exploration. We developed CLARE, a machine learning model, to predict $T_e$ using satellite, solar, and geomagnetic activity data. Instead of predicting exact $T_e$ values from a continuous range, CLARE predicts the likelihood of $T_e$ falling into 150 bins, ultimately outputting the center $T_e$ value of the most likely bin. This binned approach trades resolution for higher overall accuracy and also allows for an estimation of prediction confidence. CLARE's performance is high during calm solar periods, but lower during solar storms due to the relative rarity of solar storms in the training data.

\section{Introduction}
\label{introduction}
The Earth's plasmasphere, a torus-shaped region of cold, dense plasma surrounding Earth and extending above approximately 1000 km altitude \citep{goldstein_plasmasphere_2006}, represents a critical interface between the ionosphere and the outer magnetosphere. One of the key parameters governing the dynamics and energetics of cold plasma in this region is the electron temperature ($T_e$). This region is highly dynamic and primarily driven by fluctuations in solar activity and resulting geomagnetic disturbances.

During geomagnetically quiet periods, the plasmaspheric electron temperature is determined by photoelectron heating during the day and the nighttime thermal diffusion conducted down toward the ionosphere along the magnetic field line \citep{balan_plasmasphere_1996}.
During geomagnetically active periods, additional heating mechanisms have been suggested via cross-energy interactions with ring current ions \citep{ferradas_effects_2023}, superthermal electrons \citep{khazanov_inertia_2024} or wave-particle interactions \citep{usanova_role_2025}, which transfer energy to ambient thermal electrons in the plasmasphere. 
The capability to accurately model and predict $T_e$ is crucial not only for advancing our understanding of geospace response to solar and geomagnetic activities, but also for improving space weather forecasting capabilities.

In the plasmasphere, dynamics and energetics are highly intertwined. Plasma temperatures strongly affect the mass loading process, which are the forces exerted by the pressure gradient and ambipolar electric field. These forces are important for driving the polar wind \citep{kitamura_solar_2011}. Furthermore, the subsequent density distribution influences how the incoming heat will be redistributed along the magnetic field line \citep{khazanov_inertia_2024}. Knowledge of these temperatures is important for understanding how heating by solar radiation incident onto the ionosphere affects the density gradient, especially at low altitudes. 

Predicting the complex behavior of plasmaspheric parameters like electron temperature presents significant challenges given our current understanding of space physics. Traditional approaches often rely on physics-based models, which simulate the underlying physical processes \citep{liemohn_ring_2000}. While these models provide insights grounded in first principles, they can be computationally intensive and struggle to capture the full complexity of the system, due to the multitude of interacting drivers and incomplete physical understanding of the complex system. 

There were numerous attempts to empirically model electron temperature in the plasmasphere. \cite{evans_seasonal_1973} conducted one of the initial comprehensive studies using incoherent scatter radar data to establish early empirical models of electron temperature profiles by averaging several years of observations to characterize their seasonal and sunspot-cycle variations. Subsequently, \cite{mahajan_solar_1979} utilized in situ data from the Isis 1 and Explorer 22 satellites to directly compare electron temperatures between low and high solar activity periods, demonstrating large temperature increases driven by heat flux from the protonosphere were much stronger during periods of high solar activity. \cite{bilitza_solar_1990} later developed a model that was notable for including variations with solar activity. A different approach was taken by \cite{kutiev_analytical_2002}, who used an analytical representation of the electron temperature distribution using AKEBONO data.  \cite{titheridge_temperatures_1998} developed a computationally simple model based on an analytic solution for temperature variation along a magnetic field line, defined by the temperature and its gradient at a 400 km reference height. Building directly on Titheridge's work, \cite{webb_modifications_2003} used an expanded satellite database to refine the key parameters of that model and introduced important new modifications, such as a more realistic solar flux correction and the inclusion of seasonal variations. The \cite{truhlik_new_2012} model further expanded on previous studies by incorporating an even more extensive database of satellite measurements to provide a global empirical model with a more detailed description of solar activity variations. This model remains a key component of the latest version of the International Reference Ionosphere, IRI-2020, which is the current international standard empirical model for modeling electron temperature \citep{bilitza_international_2022}.

In recent years, machine learning (ML) has emerged as a powerful data-driven approach for tackling complex modeling tasks in many fields, including space physics. ML models excel at identifying non-linear relationships within large, high-dimensional datasets, offering a promising avenue to enhance predictive accuracy and validate physics-based empirical methods.

Early works, such as \cite{wing_kp_2005}, recognized the potential of machine learning in space weather by predicting $K_p$ indices, a measure of solar storm intensity, using neural networks. Subsequent ML research in space weather has seen considerable activity, although efforts have predominantly focused on predicting electron density ($N_e$), particularly within the ionosphere for altitudes typically below 1000 km. This focus has been largely driven by the relative abundance of data from Low Earth Orbit (LEO) satellites and ground-based instruments, when compared to data from the plasmasphere. Numerous studies have employed various neural network architectures to model $N_e$ globally or regionally \citep{chu_neural_2017, zhelavskaya_empirical_2017, li_application_2021}.
In contrast to the extensive work on electron density, the application of ML for predicting electron temperature remains comparatively underexplored due to a lack of sufficient data, especially within the higher altitude plasmasphere above 1000 km. While notable work exists, such as \cite{hu_deep_2020} who developed a deep neural network model for global topside ionospheric $T_e$ using incoherent scatter radar data, a dedicated ML model leveraging satellite measurements for predicting $T_e$ specifically within the plasmasphere has been missing. This represents a critical gap in our ability to model and forecast the state of this important region using data-driven techniques.

To address this gap, we introduce CLARE: a \textbf{Cla}ssification-based \textbf{Re}gression neural network for electron temperature prediction in the Earth's plasmasphere. CLARE is the first machine learning model specifically developed to predict electron temperature between 1000 km and 8000 km altitude using historical in-situ satellite measurements (AKEBONO data) combined with relevant solar and geomagnetic indices (NASA OMNI data). In machine learning, classification models are typically used for assigning input data into discrete categories, while regression models are used for predicting continuous numerical values. Previous machine learning models used for predicting $T_e$ have modelled the task as a regression problem as temperature is an inherently continuous physical quantity. Instead of treating $T_e$ prediction as a standard continuous regression task with a single numerical output prediction, CLARE discretizes the continuous $T_e$ range into a set of 150 bins, predicting a probability distribution over these bins. The model then outputs the center value of the most probable bin, resulting in a final numerical output.

CLARE's combined approach of a classification-based regression framework offers several key advantages for $T_e$ prediction in the plasmasphere. By discretizing the continuous output space into a finite set of bins, the model simplifies the regression task, which is particularly effective in the noisy and data-scarce conditions of the plasmasphere. This approach mitigates the effects of variance by reducing sensitivity to input noise and outliers, resulting in a 6.46\% relative performance gain compared to a traditional regression model applied to our dataset (see Table 2). The model's prediction of a probability distribution over bins also provides a built-in mechanism for uncertainty quantification, allowing for more interpretable outputs. Together, these advantages make CLARE well-suited for modeling $T_e$ in the plasmasphere given the available data.

\section{Dataset}
\label{data}

For the present study, we developed a three-dimensional electron temperature ($T_e$) model using data obtained by a thermal electron distribution (TED) instrument onboard the AKEBONO satellite \citep{abe_measurements_1990}. The AKEBONO satellite's orbit has a perigee of 300 km, an initial apogee of 10,500 km and a high 75 degree angle of inclination \citep{tsuruda_introduction_1993}. We used the data between 1000 km and 8000 km in altitude during a 10-year period from January 1st 1990 to January 1st 2001. 

The training dataset contains location-specific features and $T_e$ targets from the AKEBONO dataset and solar and geomagnetic indices (SYM-H index, AL index, and F$_{10.7}$) from the NASA OMNI dataset, following the preparation methods from \cite{chu_neural_2017}. Additionally, we added the current $K_p$ index (NASA OMNI) to give the model context of solar storm intensity. 

We split our evaluation test set into a special subset representing a known geomagnetic storm period between January 30th to February 7th of 1991 \citep{liemohn_ring_2000}, which we refer to as ``test-storm'' and a randomly sampled discontinuous period of 50,000 samples referred to as ``test-quiet''. These test sets are mutually exclusive, containing no samples in common, and are also held out of the training set.

\begin{figure}[h]
    \centering
    \includegraphics[width=1.0\textwidth]{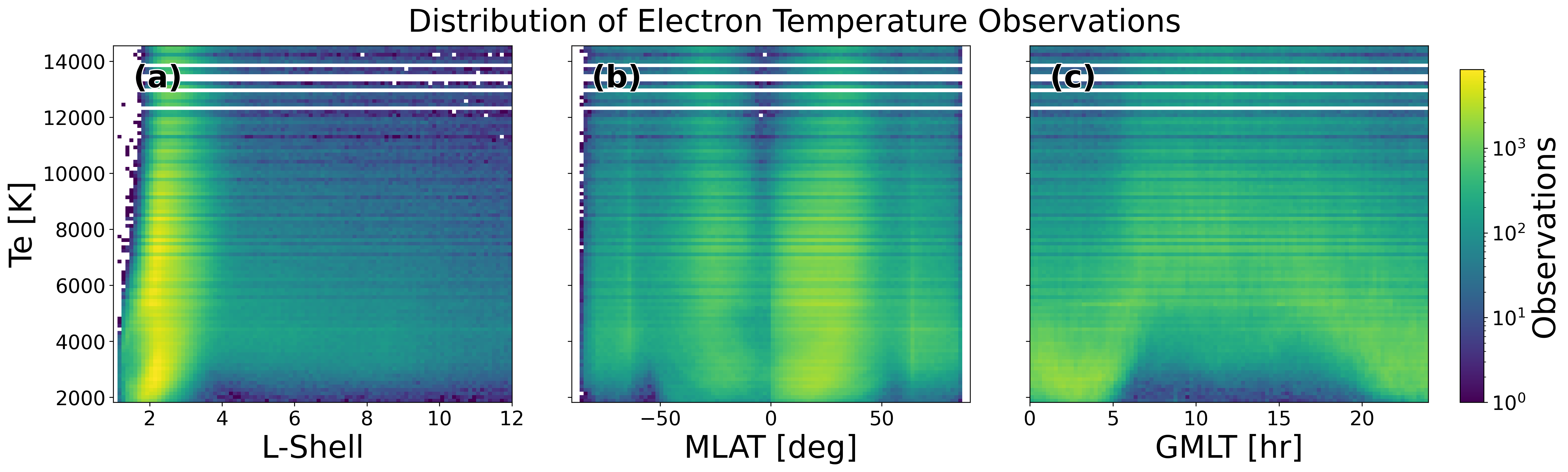}
    \caption{\centering The distributions of in situ observations of the electron temperature (Te) with respect to (a) L shell (between 1 and
12), (b) MLAT (between -90° to 90°), and (c) GMLT. The colorbar shows the number of observations in each bin on a log10 scale.}
    \label{fig:akebono_data}
\end{figure}

Figure \ref{fig:akebono_data} shows the number of in situ observations of $T_e$ with respect to L shell, MLAT, and GMLT.  Figure \ref{fig:akebono_data}a illustrates the number of observations versus $T_e$ and L shell. At lower L shells between 1.5 to 3.5, we see a concentration that extends across the $T_e$ range, signifying that most of the data is located within these L-shell values. Figure 1b shows the number of in situ observations versus $T_e$ and MLAT. The measurements are predominantly centered around the equator, with concentrated regions of measurement between -80 to -60, -40 to -5, 0 to 40, and 60 to 80 degrees. Figure 1c shows the number of in situ observations versus $T_e$ and GMLT. Higher electron temperatures are observed during the day from 5 to 20, with lower $T_e$ values from 0 to 5 and 20 to 24.

\section{Model}
\label{model}

\begin{figure}[h]
    \centering
    \includegraphics[width=1.0\textwidth]{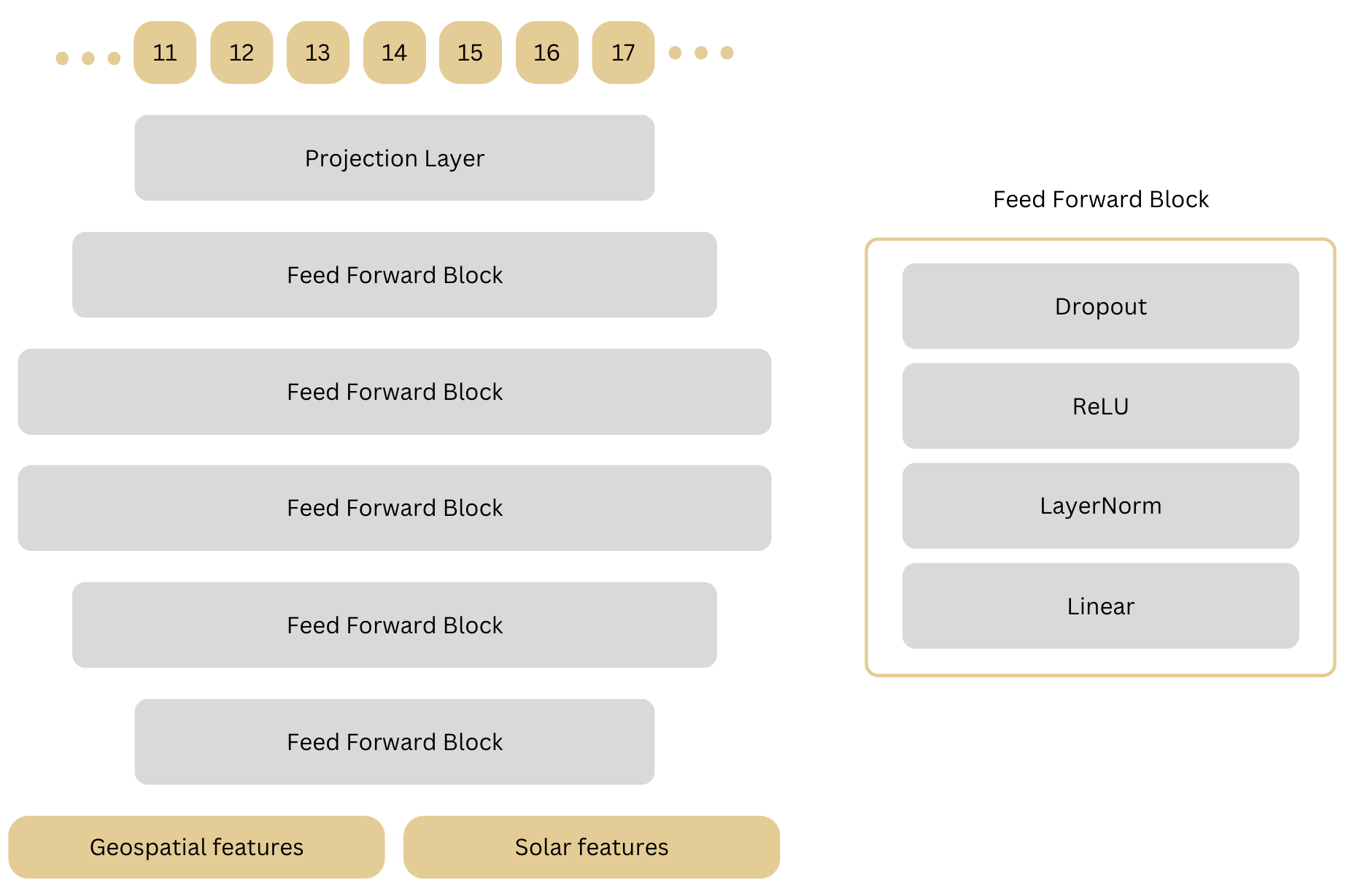}
    \caption{\centering Model architecture for CLARE. The model consists of multiple feedforward blocks, with a detailed view of a single feedforward block shown on the right. Gray components represent neural network layers, while yellow components indicate input data and output logits.}
    \label{fig:model}
\end{figure}

CLARE is an 84 million parameter 6-layer feedforward network with an initial hidden size of 2048, expanding to 8192 before projecting down to 150 neurons. Our core computational unit is the feedforward block which consists of a linear layer, followed by LayerNorm \citep{ba_layer_2016}, ReLU activation function \citep{nair_rectified_2010}, and a Dropout layer \citep{srivastava_dropout_2014}. We stack 5 of these blocks together followed by a final linear layer that projects the hidden states to 150 neurons.

We use the AdamW optimizer~\citep{loshchilov_decoupled_2019}, a standard and widely used optimizer for large neural networks, with a batch size of 512 and a cosine-decaying learning rate schedule with a linear warm-up phase, reaching a peak value of \( 0.0008 \) before decreasing to \( \frac{1}{1000} \) of the maximum learning rate. These hyperparameters are chosen empirically based on what gave the best performance in our experiments.

Continuous regression tasks are typically modeled with a single output neuron predicting a scalar value across the entire continuous output space but, inspired by work in the reinforcement learning literature \citep{tang_discretizing_2020}, we alter the continuous regression task into a classification task with a discrete action space by making the model output to 150 neurons in the final projection layer instead of a single neuron. As such, we task each neuron with predicting the probability of the electron temperature in each 100 \text{K} bin between \( [0, 15000] \, \text{K} \) where the exact bin count of 150 is determined empirically. We find this reduces the complexity of the modeling task and improves the model performance.

We then train the model using the cross-entropy loss function where the electron temperature target is the bin index corresponding to the correct bin interval.
\begin{equation}
Loss = - \sum_{i} y_i \log(p_i)
\end{equation}

During test-time, we obtain the final $T_e$ predictions on the Kelvin scale by applying the $\operatorname{argmax}$ function to the model logits and a lightweight post-processing step to shift the prediction in each bin to the center. Mathematically, if the predicted bin index is $k$, the centered electron temperature is computed as:
\begin{equation}
T_e = k \cdot \Delta T + \frac{1}{2} \Delta T,
\end{equation}
where $\Delta T = 100 \, \text{K}$. This effectively adds $50 \, \text{K}$ to center the prediction within the bin, removing bias towards either side of the bin boundaries.

Finally, our model is trained for 90 minutes on a 1 x NVIDIA V100 GPU and 8 x CPU machine through the National Center for Atmospheric Research Casper supercomputing cluster.

\newpage
\section{Results}
\label{results}
\subsection{Model Performance}
\label{results:model_performance}

\begin{figure}[!htbp]
    \centering
    \begin{subfigure}[b]{0.8\textwidth}
        \centering
        \includegraphics[width=\textwidth]{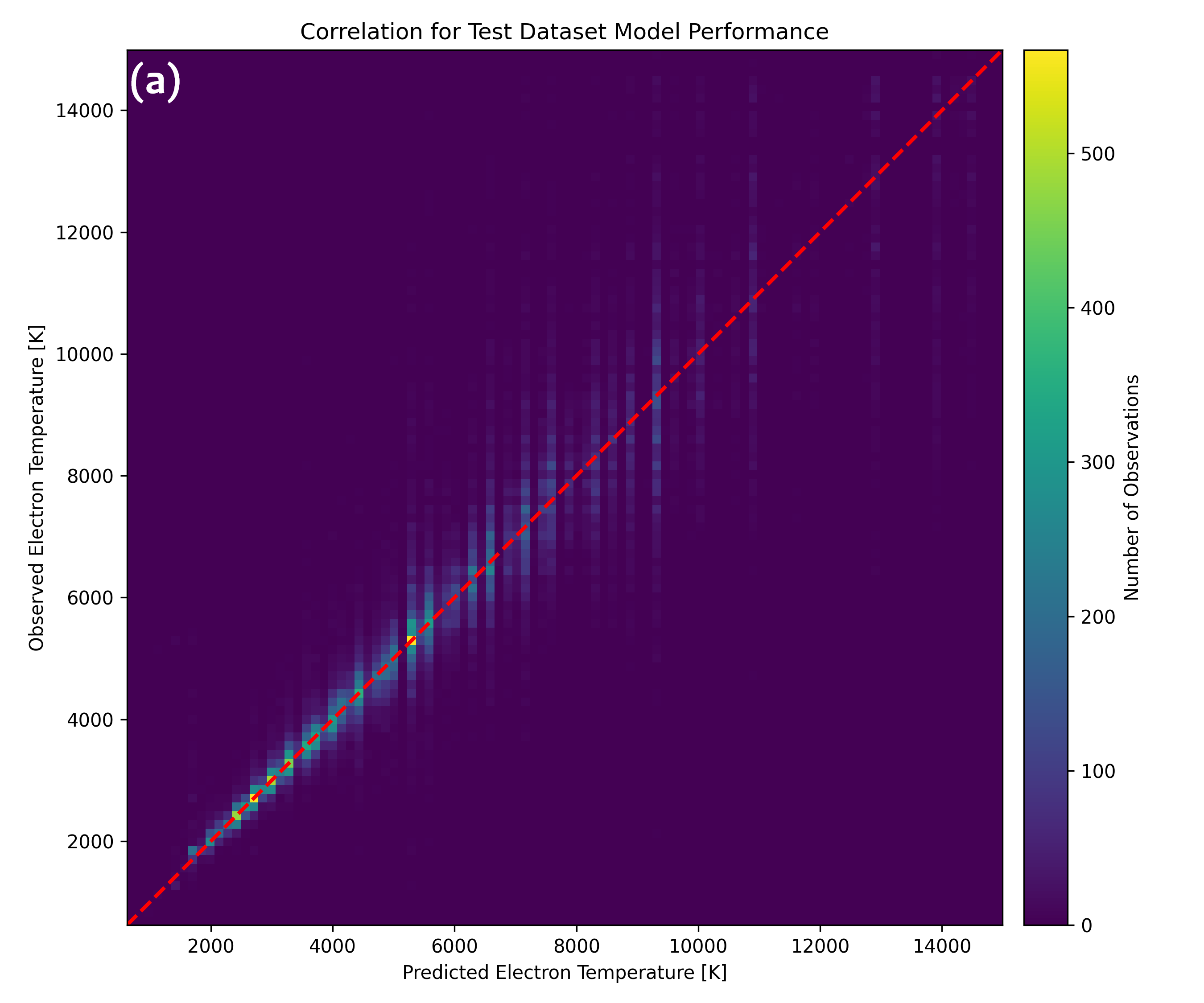}
        \label{fig:test_quiet_correlation_plot}
    \end{subfigure}
    \begin{subfigure}[b]{0.8\textwidth}
        \centering
        \includegraphics[width=\textwidth]{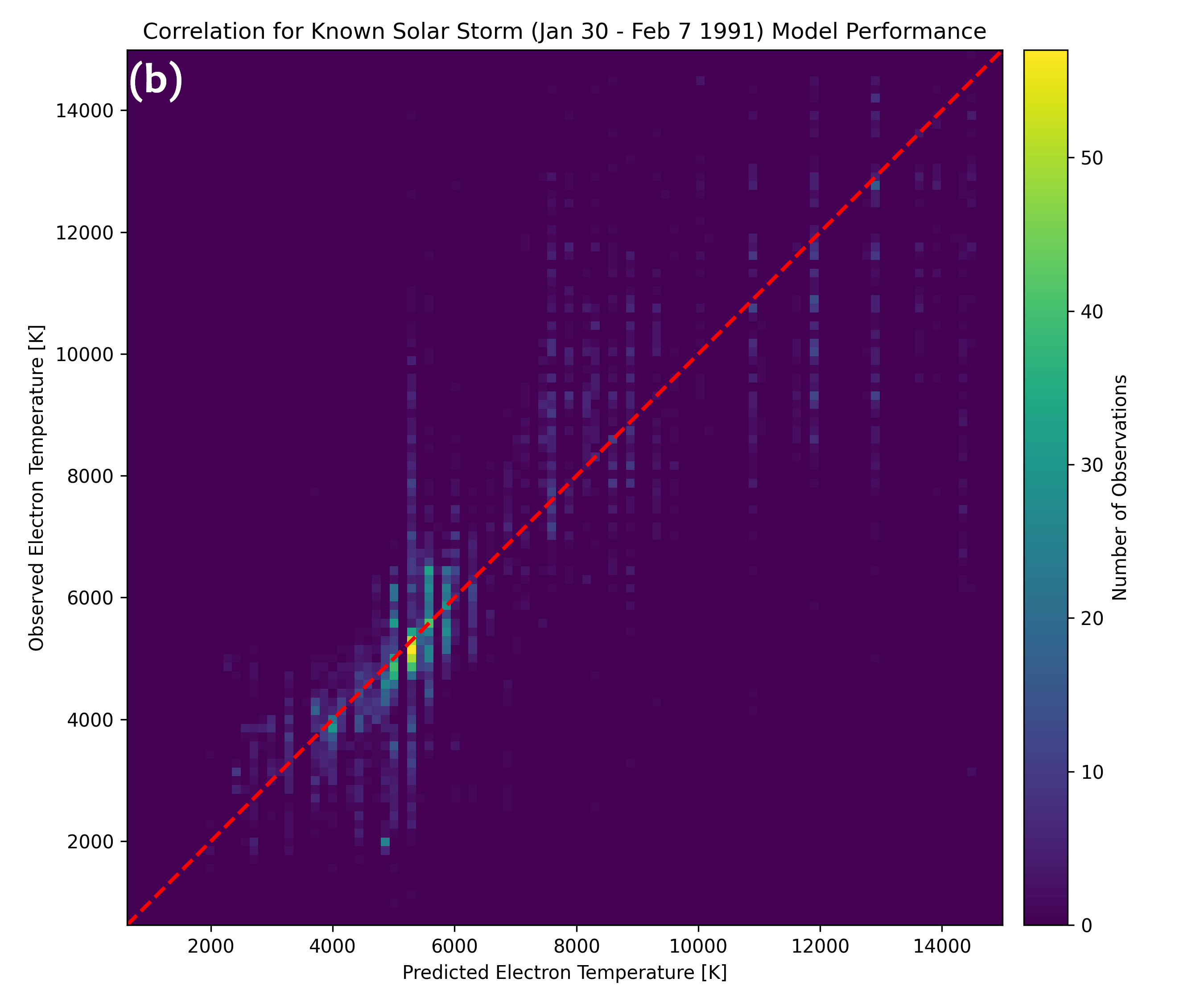}
        \label{fig:test_storm_correlation_plot}
    \end{subfigure}
    
    \caption{Correlation between observed and predicted electron temperature on the (a) test dataset of 50,000 randomly held out samples and (b) the held out known solar storm period (January 30th through February 7th, 1991).}
    \label{fig:combined_correlation_plots}
\end{figure}
\FloatBarrier

Figure \ref{fig:combined_correlation_plots}(a) demonstrates the high performance of the CLARE model on a representative random sample of 50,000 points held out from the training data set. The spread of the number of observations increases as the electron temperature increases, demonstrating that model performance decreases as the electron temperature increases. This is likely due to the greater degree of variability in the training data for higher temperature, which is associated with increases in altitude, as Figure 1(a) shows. The impacts on model performance during high solar storm periods can be seen in Figure \ref{fig:combined_correlation_plots}(b). This plot demonstrates the lower performance of the CLARE model on a known solar storm period from January 30th to February 7th of 1991. This solar storm period was held out of the training and test datasets so it could be used to stress test the CLARE model's performance during periods of high solar activity.

\begin{table}[htbp]
    \centering
    \caption{Results between the Titheridge models and CLARE on quiet solar activity and solar storm test sets. Metrics include accuracy within 10\% of the observed value (higher is better), coefficient of determination $R^2$ (higher is better) and root mean square error RMSE (lower is better). Bolded results indicate the best result in the column.}
    \begin{tabular}{lcccccc}
        \toprule
        \multirow{2}{*}{Model} & \multicolumn{3}{c}{test-quiet} & \multicolumn{3}{c}{test-storm} \\
        \cmidrule(lr){2-4} \cmidrule(lr){5-7}
        & Accuracy & R$^2$ & RMSE & Accuracy & R$^2$ & RMSE \\
        \midrule
        \\[-0.8em]
        Titheridge & 11.90\% & -0.7196 & 3615.9753 & 4.96\% & -1.2208 &  4304.8868 \\[0.5em]
        Titheridge-IRI & 13.49\% & -0.4455 & 3335.7404 & 12.26\% & -0.3274 & 3170.2374 \\[0.5em]        
        CLARE & \textbf{69.67\%} & \textbf{0.7829} & \textbf{1292.6217} & \textbf{46.17\%} & \textbf{0.6647} & \textbf{1593.4668} \\[0.5em]
        \bottomrule
    \end{tabular}
    \label{tab:model_comparison}
\end{table}

\begin{figure}[h]
    \centering
    \includegraphics[width=0.8\textwidth]{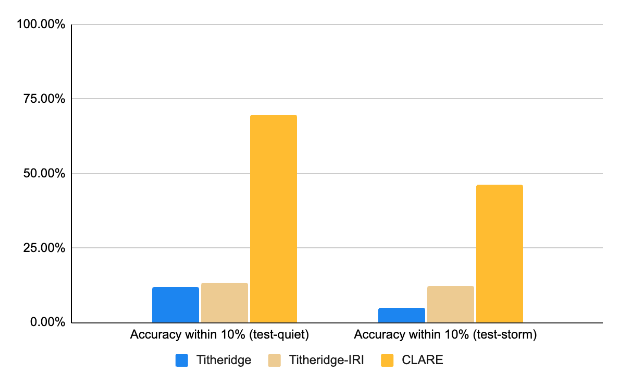}
    \caption{\centering Prediction accuracy of CLARE within 10\% of observed electron temperature against Titheridge and Titheridge-IRI baselines on test-quiet and test-storm sets.}
    \label{fig:results}
\end{figure}

Figure \ref{fig:results} and Table \ref{tab:model_comparison} demonstrate CLARE achieves 69.67\% accuracy to within 10\% tolerance of observed values during quiet solar activity periods (test-quiet) and 46.17\% accuracy during a known solar storm (test-storm), representing a significant increase above the 13.49\%/12.26\% (Titheridge-IRI quiet/storm) and 11.90\%/4.96\% (Titheridge quiet/storm) accuracy achieved with predictions from the Titheridge baselines. The coefficient of determination, R$^2$, is a measure of the correlation between the dependent variable (model predictions) to movements in the independent variable (observed $T_e$ values) where 0 and negative values represent low or negative correlation and 1 represent strong correlation. CLARE achieves an R$^2$ value of 0.7829 on the test-quiet test set and 0.6647 on the test-storm test set, demonstrating the strong correlation of its predictions with observed values.

Root mean squared error (RMSE) is a statistical metric used to measure the accuracy of a model on a regression task where higher scores represent on average greater error on the dataset and vice versa. Our model also achieves the lowest root mean squared error (RMSE) among all baselines at 1292.6217/1593.4668 (quiet/storm).

To calculate the original Titheridge baseline, we use the empirical model proposed by Equation 13 in \cite{titheridge_temperatures_1998}\footnote{The original Titheridge paper proposes many adjustments to the core equation to account for various edge cases such as diurnal variations, but we do not implement these to reduce complexity.} and apply the log10 modification identified as an error in the original equation in \cite{webb_modifications_2003}. To ensure equality in the evaluation dataset in all our tested models, we expand the time intervals for the day coefficients to 0900 to 2100 and night coefficients to 2100 to 0900.\footnote{More information on the Titheridge equations and additional analysis following the diurnal cycles in the original paper is shared in \ref{model-performance-diurnal} and \ref{titheridge-model}.}

For the Titheridge-IRI baseline, we update the original Titheridge equation by building on top of it with the empirically derived IRI-2020 model \citep{bilitza_international_2022} to obtain reference $T_0$ values for each time step. The value of $G_0$ is determined using the least squares fit method outlined in the original paper. A known limitation of the IRI model is performance degradation above altitudes of 2000 km so we fit our $G_0$ values only on altitude values below 2000 km and then extrapolate to altitudes beyond 2000 km (where most of the EXOS-D measurements lie).\footnote{The calculation of Titheridge-IRI is expanded upon in \ref{titheridge-iri-calculation}.}

\subsection{Uncertainty Estimation Capabilities}
\label{results:uncertainty_estimation_capabilities}

A unique feature of CLARE is its ability to provide uncertainty information from its predictions by interpreting the shape of the probability mass function across all histogram bins. Probabilities are obtained by applying the softmax function on model logits:

\begin{equation}
p_i = \frac{e^{z_i}}{\sum_{j=1}^{N} e^{z_j}}
\end{equation}

where $p_i$ is the probability assigned to bin $i$, $z_i$ is the logit value for bin $i$, and $N$ is the total number of histogram bins.

\begin{figure}[h]
    \centering
    \includegraphics[width=1.0\textwidth]{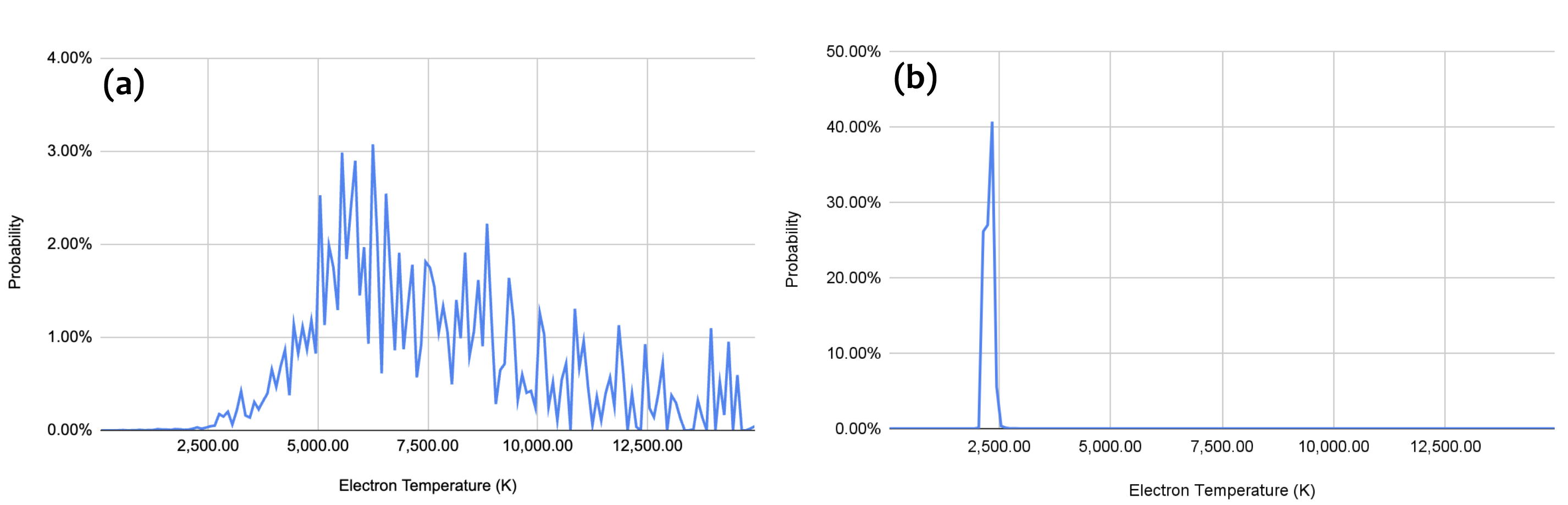}
    \caption{\centering (a) Low confidence probability distribution during high solar storm activity time instant compared to (b) high confidence probability distribution during quiet solar storm activity time instant.}
    \label{fig:entropy}
\end{figure}

As seen in Figure~\ref{fig:entropy}, sharp probability distributions with high maximum values are indicative of more confident predictions, while wide distributions with low maximum values indicate greater prediction uncertainty. We can interpret these distributions as the model providing confidence information which can be used in conditional logic during test-time.

\subsection{Impact of Discrete and Continuous Training Objective}
\label{results:impact_of_predicting_bin_probabilities}

\begin{table}[htbp]
    \centering
    \caption{Ablation results for CLARE variants with different training objectives. Bolded results indicate the best result in the column.}    
    \begin{tabular}{lcccc}
        \toprule
        \multirow{2}{*}{Model} & \multicolumn{2}{c}{test-quiet} & \multicolumn{2}{c}{test-storm} \\
        \cmidrule(lr){2-3} \cmidrule(lr){4-5}
        & Accuracy & RMSE & Accuracy & RMSE \\
        \midrule
        \\[-0.8em]
        CLARE & \textbf{69.67\%} & 1292.6217 & \textbf{46.17\%} &  1593.4668 \\[0.5em]
        CLARE-continuous & 65.17\%  & \textbf{1132.0842} & 45.95\% & \textbf{1340.6564} \\[0.5em]
        \bottomrule
    \end{tabular}
    \label{tab:loss_ablation}
\end{table}

To determine the value of the discrete bin prediction objective, we perform an ablation, testing CLARE against a model variant tasked with outputting a single scalar prediction value (CLARE-continuous). All model training hyperparameters are held constant except the number of output neurons is reduced from 150 to 1 and the training objective is changed from cross entropy loss to mean squared error loss. The CLARE-continuous architecture mirrors what is standard in most of the $T_e$ prediction literature.

We observe that the binned variant performs better 6.46\% relative on the quiet solar activity periods and 0.48\% relative on the solar storm period. We also observe that the continuous variant has a lower RMSE value across both test sets.

Table \ref{tab:loss_ablation} reveals that the cross entropy training objective is better suited for raw performance accuracy but MSE loss favors reducing the variance of outliers.

\subsection{Impact of Spatial and Solar Features}
\label{results:impact_of_input_features}

\begin{table}[htbp]
    \centering
    \caption{Ablation results for CLARE trained on different datasets. Bolded results indicate the best result in the column.}
    \begin{tabular}{lcccccc}
        \toprule
        \multirow{2}{*}{Model} & \multicolumn{3}{c}{test-quiet} & \multicolumn{3}{c}{test-storm} \\
        \cmidrule(lr){2-4} \cmidrule(lr){5-7}
        & Accuracy & R$^2$ & RMSE  & Accuracy & R$^2$ & RMSE \\
        \midrule
        \\[-0.8em]
        CLARE & \textbf{69.67\%} & \textbf{0.7829} & \textbf{1292.6217} & 46.17\% & \textbf{0.6647} & \textbf{1593.4668} \\[0.5em]
        CLARE-solar & 66.69\% & 0.7504 & 1385.9779 & 13.79\% & -0.5679 & 3445.5452 \\[0.5em]
        CLARE-spatial & 58.38\% & 0.6492 & 1643.2788  & \textbf{47.38\%} & 0.6254 & 1684.1300 \\[0.5em]
        \bottomrule
    \end{tabular}
    \label{tab:dataset_ablation}
\end{table}

We also perform ablations on the input features to determine the relative impact of the spatial and solar indices to the modeling task by training additional models with only the solar indices data from the NASA OMNI dataset (CLARE-solar) and only the spatial data from the JAXA AKEBONO dataset (CLARE-spatial).

As expected, CLARE trained with both solar and spatial data performs overall the best on both test sets. We find that, on our quiet solar activity periods, training on only the solar features incurs a 4.27\% relative degradation compared to a 16.2\% relative degradation when removing solar features from the dataset. This suggests that CLARE is deriving significantly stronger prediction signal from solar features than spatial features during quiet solar activity periods.

Interestingly, CLARE-spatial (trained only on spatial features) has the best overall accuracy on the chaotic storm test set which suggests that the solar storm prediction task is highly correlated with the state of the spacecraft as opposed to the solar environment in general.

\section{Event Analysis}
\label{Event Analysis}

\begin{figure}[h]
    \centering
    \includegraphics[width=1.0\textwidth]{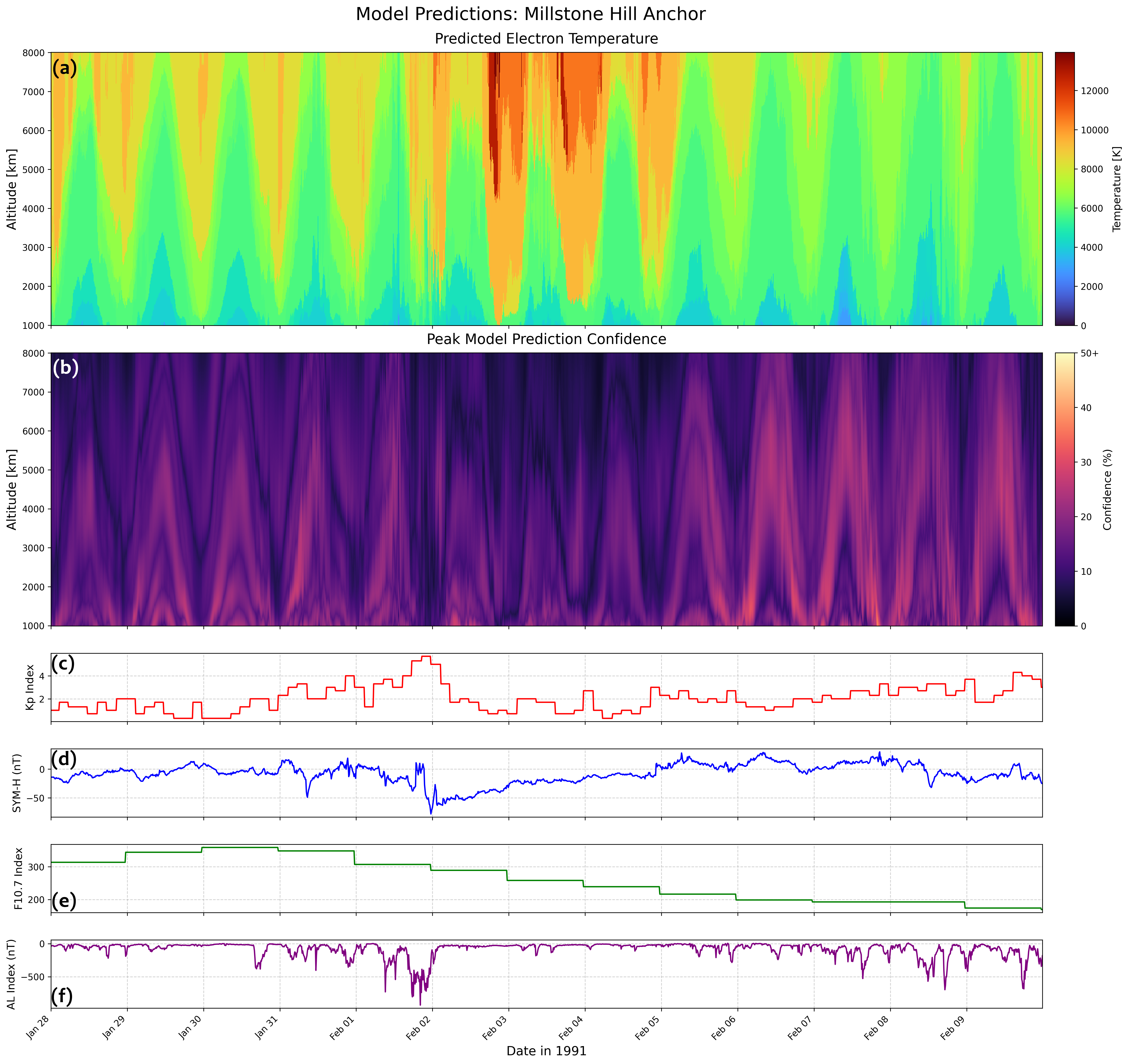}
    \caption{\centering Model predicted $T_e$ values during a moderate intensity solar storm between January 30th and February 7th, 1991, for altitudes between 1000 km and 8000 km fixed above Millstone Hill Observatory (geographic latitude 43$^\circ$, longitude 289$^\circ$). The plot contains (a) the predicted $T_e$ value, (b) the peak model prediction confidence for the corresponding predicted $T_e$ value, and solar features from the NASA OMNI dataset, including (c) Kp index, (d) SYM-H, (e) F10.7 index, and (f) AL index.}
    \label{fig:event_analysis}
\end{figure}

 A medium intensity solar storm (SYM-H peak of approximately -100) occurred between January 30th and February 7th 1991. Thus, we hold out this time interval from both the training and test datasets and use it to verify the model's performance in predicting the electron temperature response to a geomagnetic storm. Figure~\ref{fig:event_analysis} shows the predictions of the model during this time period.
Before the storm that commenced on February 1, the electron temperature exhibited a diurnal variation as observed above the Millstone Hill Observatory (geographic latitude 43$^\circ$, longitude 289$^\circ$) with the reference frame rotating with the Earth. After the storm started, the model predicted higher $T_e$ values, with the enhancement extending to lower altitudes and gradually decaying as the storm recovered on February 5, demonstrating a physically reasonable response. Previous studies suggest that the energy source for heating plasmaspheric electrons originates in the equatorial magnetosphere \citep{kozyra_high-altitude_1997, balan_plasmasphere_1996}, either through Coulomb collisions with ring current ions \citep{ferradas_effects_2023} or through wave-particle interactions to transfer energy between hot and thermal populations \citep{kozyra_high-altitude_1997, cornwall_turbulent_1970, hasegawa_anomalous_1978}.

During this period, 46.17\% of the model's predictions are within 10\% of the actual ground-truth value. Solar storm periods are rare events, comprising a fraction of the total dataset, and contain highly variable solar indices and observed ground-truth $T_e$ values, sometimes with $T_e$ fluctuating by thousands of Kelvin within minutes. Thus, we consider this level of performance to be strong, as the model learns the underlying patterns of the space environment with a limited number of examples and unstable input features and output targets.

\section{Discussion}
\label{Discussion}
In this paper, we build upon techniques in the electron density literature \citep{chu_neural_2017, zhelavskaya_empirical_2017, li_application_2021} and apply it to the electron temperature prediction task. CLARE introduces a more performant modification to the traditional feedforward model architecture and scales the dataset size and model parameter count to roughly an order of magnitude larger than most research in the field. Our research is in line with the broader machine learning community as it trends towards larger models trained on more data and compute \citep{kaplan_scaling_2020}.

Uncertainty quantification is important in real-world use cases to know whether or not to trust the predictions out of inherently stochastic systems. Our approach to determining uncertainty differs from others in the literature who achieve this capability by adding additional computation modules \citep{zhan_quantifying_2024,camporeale_accrue_2021}. We opt instead to jointly train this capability alongside the main $T_e$ prediction task by tasking the model to predict a probability mass function instead of a scalar value. Our approach sits alongside other model-embedded approaches like Bayesian neural networks \citep{siddique_survey_2022} and can be considered a simpler alternative.

Large amounts of high-quality data aligned with the physics of the real world are extremely important to any space physics machine learning model as they try to generalize to the real world from the distribution of their training data. As solar storm time periods are significantly rarer than quiet solar periods, representing only 0.74\% of the dataset\footnote{Roughly estimated as the percentage of samples that have SYM-H values less than -100.}, we have a data imbalance that makes it extremely difficult for any machine learning model to learn the storm environment. As a result, our model tends to predict smooth distributions biased towards the mean without fully capturing the complexity of short-term variations. This is a key limitation of the machine learning technique that can be solved by scaling model and dataset size to give the model greater capacity to fit to these complex relationships or more sophisticated modeling techniques such as weighted losses explored in \cite{chu_imbalanced_2025}.

Furthermore, we believe merging established physics-based approaches with  machine learning techniques could produce even better results. Physics-based approaches represent a distillation of human understanding of the underlying physics phenomena and may reveal to the model patterns not fully captured in empirical measurements. Empirical satellite measurements of the space environment are frequently plagued with noise arising from various sources such as instrumentation tuning, orbital dynamics and artefacts of the space environment. We believe it would be worthwhile to explore synthetic data generation from machine learning models to inform physics-based models or physics-based predictions used as inputs into machine learning models to push the performance of space weather prediction models even further.

\section{Conclusion}
\label{conclusion}

We present CLARE, the first machine learning model specifically designed for predicting electron temperature in the Earth's plasmasphere between 1000\,km and 8000\,km altitude. By reframing the continuous $T_e$ regression task as a classification problem over 150 discrete bins, CLARE achieves 69.67\% accuracy within 10\% relative tolerance on quiet solar activity periods, substantially outperforming the Titheridge (11.90\%) and Titheridge-IRI (13.49\%) empirical baselines. On a held-out geomagnetic storm period from January 30th to February 7th, 1991, CLARE maintains 46.17\% accuracy despite storm conditions comprising only 0.74\% of the training data.

Ablation experiments reveal that the classification-based training objective yields a 6.46\% relative accuracy gain over an equivalent continuous regression architecture, while simultaneously providing built-in uncertainty quantification through its predicted probability distributions.

A key limitation of this work is the model's reduced performance during geomagnetic storms, driven by the inherent scarcity of storm-time observations in the training data. Future work could address this through data augmentation strategies, weighted loss functions, or the integration of physics-based model outputs as additional input features to provide the model with physically grounded priors not fully captured in empirical measurements alone. More broadly, CLARE demonstrates that the plasmasphere, long underserved by data-driven methods, is amenable to high-accuracy machine learning modeling using publicly available data, opening the door to improved space weather forecasting in this critical region.

\section*{Open Research}
\label{Open Research}

The code is available at \url{https://github.com/blakedehaas/clare}

AKEBONO (EXOS-D) Satellite data is available at \url{https://darts.isas.jaxa.jp/stp/akebono/TED.html}

Solar indices from NASA OMNI is available at \url{https://omniweb.gsfc.nasa.gov/}

IRI 2020 is available at \url{https://github.com/MIST-Experiment/iricore}

\acknowledgments

We would like to thank the National Aeronautics and Space Administration (NASA) for funding this research and the National Center for Atmospheric Research for providing supercomputer resources.

Xiangning Chu would like to thank grant NASA 80NSSC22K1023, 80NSSC23K0096, 80NSSC24K1112, NSF grant AGS-2247255, and AFOSR YIP FA9550-23-1-0359.

\appendix
\label{Appendix}
\section{Titheridge and Titheridge-IRI verification}
\label{titheridge-iri-baseline}
\begin{figure}[!htb]
    \centering
    \includegraphics[width=1.0\textwidth]{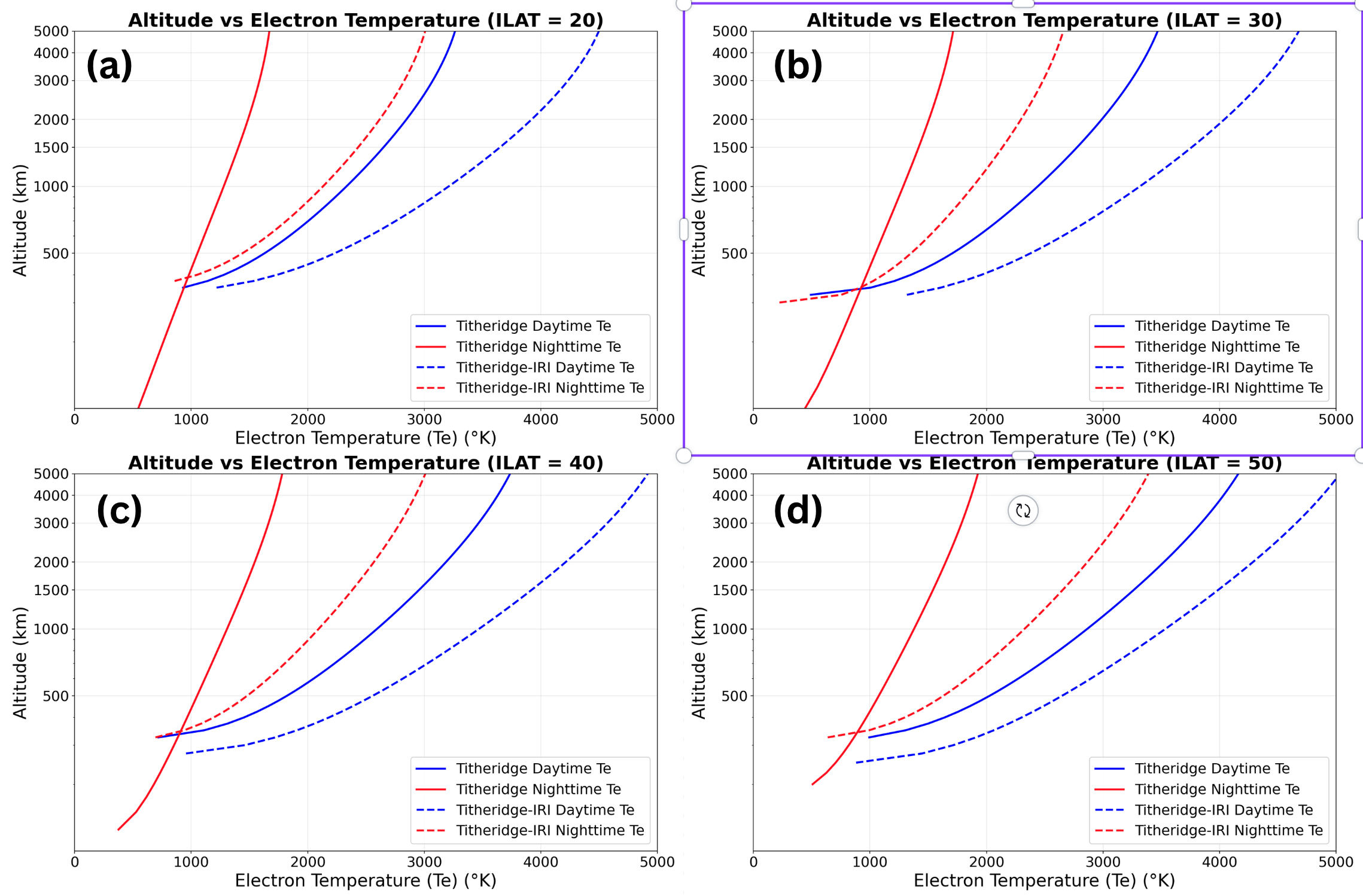}
    \caption{Electron temperature to altitude comparisons for Titheridge and Titheridge-IRI for nighttime and daytime at (a) ILAT = 20, (b) ILAT = 30, (c) ILAT = 40, and (d) ILAT = 50.}
    \label{fig:webb_replication_fig6}
\end{figure}

To verify integrity of the Titheridge baselines, we replicate Figure 6 of the  \cite{webb_modifications_2003} paper on our quiet solar activity test set as shown in our own Figure \ref{fig:webb_replication_fig6}, showing similar curve shapes and boundaries for all 4 latitudes considered. Note that exact replication is not expected as our dataset is different from the authors. We consider measurements taken between 0900 to 1600 to be daytime and 2100 to 0400 to be nighttime according to the original paper.

\section{Titheridge Model}
\label{titheridge-model}

We use the following equation from the \cite{titheridge_temperatures_1998} paper to calculate our baseline
\begin{equation}
T_e(h) = T_0 \left[ 1 + B(h) \frac{G_0}{T_0} \left( \frac{h_{\text{eq}} - h_0}{R_0^2} - \frac{h_{\text{eq}} - h}{R_h^2} \right) \right]^{2/7}
\label{eq:3}
\end{equation}

For the $G_0$ and $T_0$ free parameter values, we use Equation \ref{eq:4} and the coefficients, both proposed in the original paper.

\begin{equation}
X(s_L) = \frac{(a_0 + a_1 s_L + a_2 s_L^2)}{(1 + a_3 s_L + a_4 s_L^2)}
\label{eq:4}
\end{equation}

\section{Model performance on diurnal filtered dataset}
\label{model-performance-diurnal}
The day and night coefficients associated with Equation \ref{eq:3} of \cite{titheridge_temperatures_1998} is typically tied to the time intervals 0900 - 1600 (day) and 2100 - 0400 (night) so we also provide performance metrics across the Titheridge baselines and the CLARE model for the filtered test sets where we remove all time steps outside of those ranges.

Table \ref{tab:model_comparison_diurnal} shows similar results trends to the models evaluated on the unfiltered dataset in Table {\ref{tab:model_comparison}.

\begin{table}[htbp]
    \caption{Results between the Titheridge models and CLARE on quiet solar activity and solar storm test sets filtered on the day-night times of the original Titheridge paper. Bolded results indicate the best result in the column.}
    \centering
    \begin{tabular}{lcccccc}
        \toprule
        \multirow{2}{*}{Model} & \multicolumn{3}{c}{test-quiet} & \multicolumn{3}{c}{test-storm} \\
        \cmidrule(lr){2-4} \cmidrule(lr){5-7}
        & Accuracy & R$^2$ & RMSE & Accuracy & R$^2$ & RMSE \\
        \midrule
        \\[-0.8em]
        Titheridge & 12.34\% & -0.7115 & 3610.7134 & 3.46\% & -1.5314 &  3918.1794 \\[0.5em]
        Titheridge-IRI & 13.55\% & -0.5086 & 3389.9854 & 12.53\% & -0.7376 & 3246.2239 \\[0.5em]        
        CLARE & \textbf{69.84\%} & \textbf{0.7936} & \textbf{1253.8517} & \textbf{22.91\%} & \textbf{-0.6071} & \textbf{3121.9749} \\[0.5em]
        \bottomrule
    \end{tabular}
    \label{tab:model_comparison_diurnal}
\end{table}

\section{Titheridge-IRI methodology}
\label{titheridge-iri-calculation}

As the original Titheridge coefficients were calculated over 20 years ago, we also improve upon this baseline by incorporating the most recent IRI model to calculate new $G_0$ and $T_0$ values.

For each timestep in our test set, we calculate $T_0$ as the IRI-2020 model's prediction for the electron temperature at the 400 kilometer altitude. We use the date time value in UTC and the geodetic latitude and longitude from the AKEBONO dataset as inputs to the IRI model.

For the $G_0$ value, we run a linear search through all values between -20 to +50 at 0.1 increments and for each $G_0$ value, we calculate the predicted $T_e$ values according to Equation \ref{eq:3} for every 100 kilometer increment between 400 kilometers to 2000 kilometers altitude. We choose to stop at 2000 kilometers because that is the upper altitude limit in which the IRI-2020 model is accurate. Then we compute the squared error between the predicted $T_e$ values and the actual values provided by the IRI-2020 model for each $G_0$ value and choose the $G_0$ value that minimizes this value.

In effect, for each time step, we select the $G_0$ value that adjusts the gradient of the Titheridge Equation \ref{eq:3} to most closely resemble the IRI-2020 model between the altitudes of 400 kilometers and 2000 kilometers (where IRI-2020 is most accurate), then use that $G_0$ to extrapolate beyond 2000 kilometers altitude for our baseline.

\newpage
\bibliography{zotero}

\begin{thebibliography}{}

\bibitem [\protect \citeauthoryear {%
Abe%
, Okuzawa%
, Oyama%
, Amemiya%
\BCBL {}\ \BBA {} Watanabe%
}{%
Abe%
\ \protect \BOthers {.}}{%
{\protect \APACyear {1990}}%
}]{%
abe_measurements_1990}
\APACinsertmetastar {%
abe_measurements_1990}%
\begin{APACrefauthors}%
Abe, T.%
, Okuzawa, T.%
, Oyama, K\BHBI I.%
, Amemiya, H.%
\BCBL {}\ \BBA {} Watanabe, S.%
\end{APACrefauthors}%
\unskip\
\newblock
\APACrefYearMonthDay{1990}{{\APACmonth{01}}}{}.
\newblock
{\BBOQ}\APACrefatitle {Measurements of temperature and velocity distribution of thermal electrons by the {Akebono} ({EXOS}-{D}) satellite - {Experimental} setup and preliminary results} {Measurements of temperature and velocity distribution of thermal electrons by the {Akebono} ({EXOS}-{D}) satellite - {Experimental} setup and preliminary results}.{\BBCQ}
\newblock
\APACjournalVolNumPages{Journal of Geomagnetism and Geoelectricity}{42}{}{537--554}.
\newblock
\begin{APACrefURL} [{2025-08-14}]\url{https://ui.adsabs.harvard.edu/abs/1990JGG....42..537A} \end{APACrefURL}
\newblock
\APACrefnote{ADS Bibcode: 1990JGG....42..537A}
\newblock
\begin{APACrefDOI} \doi{doi:10.5636/jgg.42.537} \end{APACrefDOI}
\PrintBackRefs{\CurrentBib}

\bibitem [\protect \citeauthoryear {%
Ba%
, Kiros%
\BCBL {}\ \BBA {} Hinton%
}{%
Ba%
\ \protect \BOthers {.}}{%
{\protect \APACyear {2016}}%
}]{%
ba_layer_2016}
\APACinsertmetastar {%
ba_layer_2016}%
\begin{APACrefauthors}%
Ba, J\BPBI L.%
, Kiros, J\BPBI R.%
\BCBL {}\ \BBA {} Hinton, G\BPBI E.%
\end{APACrefauthors}%
\unskip\
\newblock
\APACrefYearMonthDay{2016}{{\APACmonth{07}}}{}.
\newblock
\APACrefbtitle {Layer {Normalization}.} {Layer {Normalization}.}
\newblock
\APACaddressPublisher{}{arXiv}.
\newblock
\begin{APACrefURL} [{2025-05-22}]\url{http://arxiv.org/abs/1607.06450} \end{APACrefURL}
\newblock
\APACrefnote{arXiv:1607.06450 [stat]}
\newblock
\begin{APACrefDOI} \doi{10.48550/arXiv.1607.06450} \end{APACrefDOI}
\PrintBackRefs{\CurrentBib}

\bibitem [\protect \citeauthoryear {%
Balan%
, Oyama%
, Bailey%
\BCBL {}\ \BBA {} Abe%
}{%
Balan%
\ \protect \BOthers {.}}{%
{\protect \APACyear {1996}}%
}]{%
balan_plasmasphere_1996}
\APACinsertmetastar {%
balan_plasmasphere_1996}%
\begin{APACrefauthors}%
Balan, N.%
, Oyama, K\BHBI I.%
, Bailey, G\BPBI J.%
\BCBL {}\ \BBA {} Abe, T.%
\end{APACrefauthors}%
\unskip\
\newblock
\APACrefYearMonthDay{1996}{}{}.
\newblock
{\BBOQ}\APACrefatitle {Plasmasphere electron temperature profiles and the effects of photoelectron trapping and an equatorial high-altitude heat source} {Plasmasphere electron temperature profiles and the effects of photoelectron trapping and an equatorial high-altitude heat source}.{\BBCQ}
\newblock
\APACjournalVolNumPages{Journal of Geophysical Research: Space Physics}{101}{A10}{21689--21696}.
\newblock
\begin{APACrefURL} \url{https://agupubs.onlinelibrary.wiley.com/doi/abs/10.1029/96JA01798} \end{APACrefURL}
\newblock
\begin{APACrefDOI} \doi{https://doi.org/10.1029/96JA01798} \end{APACrefDOI}
\PrintBackRefs{\CurrentBib}

\bibitem [\protect \citeauthoryear {%
Bilitza%
\ \BBA {} Hoegy%
}{%
Bilitza%
\ \BBA {} Hoegy%
}{%
{\protect \APACyear {1990}}%
}]{%
bilitza_solar_1990}
\APACinsertmetastar {%
bilitza_solar_1990}%
\begin{APACrefauthors}%
Bilitza, D.%
\BCBT {}\ \BBA {} Hoegy, W\BPBI R.%
\end{APACrefauthors}%
\unskip\
\newblock
\APACrefYearMonthDay{1990}{{\APACmonth{01}}}{}.
\newblock
{\BBOQ}\APACrefatitle {Solar activity variation of ionospheric plasma temperatures} {Solar activity variation of ionospheric plasma temperatures}.{\BBCQ}
\newblock
\APACjournalVolNumPages{Advances in Space Research}{10}{8}{81--90}.
\newblock
\begin{APACrefURL} [{2025-08-13}]\url{https://www.sciencedirect.com/science/article/pii/027311779090190B} \end{APACrefURL}
\newblock
\begin{APACrefDOI} \doi{10.1016/0273-1177(90)90190-B} \end{APACrefDOI}
\PrintBackRefs{\CurrentBib}

\bibitem [\protect \citeauthoryear {%
Bilitza%
\ \protect \BOthers {.}}{%
Bilitza%
\ \protect \BOthers {.}}{%
{\protect \APACyear {2022}}%
}]{%
bilitza_international_2022}
\APACinsertmetastar {%
bilitza_international_2022}%
\begin{APACrefauthors}%
Bilitza, D.%
, Pezzopane, M.%
, Truhlik, V.%
, Altadill, D.%
, Reinisch, B\BPBI W.%
\BCBL {}\ \BBA {} Pignalberi, A.%
\end{APACrefauthors}%
\unskip\
\newblock
\APACrefYearMonthDay{2022}{}{}.
\newblock
{\BBOQ}\APACrefatitle {The {International} {Reference} {Ionosphere} {Model}: {A} {Review} and {Description} of an {Ionospheric} {Benchmark}} {The {International} {Reference} {Ionosphere} {Model}: {A} {Review} and {Description} of an {Ionospheric} {Benchmark}}.{\BBCQ}
\newblock
\APACjournalVolNumPages{Reviews of Geophysics}{60}{4}{e2022RG000792}.
\newblock
\begin{APACrefURL} [{2025-05-22}]\url{https://onlinelibrary.wiley.com/doi/abs/10.1029/2022RG000792} \end{APACrefURL}
\newblock
\begin{APACrefDOI} \doi{10.1029/2022RG000792} \end{APACrefDOI}
\PrintBackRefs{\CurrentBib}

\bibitem [\protect \citeauthoryear {%
Camporeale%
\ \BBA {} Carè%
}{%
Camporeale%
\ \BBA {} Carè%
}{%
{\protect \APACyear {2021}}%
}]{%
camporeale_accrue_2021}
\APACinsertmetastar {%
camporeale_accrue_2021}%
\begin{APACrefauthors}%
Camporeale, E.%
\BCBT {}\ \BBA {} Carè, A.%
\end{APACrefauthors}%
\unskip\
\newblock
\APACrefYearMonthDay{2021}{}{}.
\newblock
{\BBOQ}\APACrefatitle {{ACCRUE}: {ACCURATE} {AND} {RELIABLE} {UNCERTAINTY} {ESTIMATE} {IN} {DETERMINISTIC} {MODELS}} {{ACCRUE}: {ACCURATE} {AND} {RELIABLE} {UNCERTAINTY} {ESTIMATE} {IN} {DETERMINISTIC} {MODELS}}.{\BBCQ}
\newblock
\APACjournalVolNumPages{International Journal for Uncertainty Quantification}{11}{4}{}.
\newblock
\begin{APACrefURL} [{2025-08-14}]\url{https://www.dl.begellhouse.com/journals/52034eb04b657aea,3ec0b84376cff3d2,1801e97431c5911b.html} \end{APACrefURL}
\newblock
\begin{APACrefDOI} \doi{10.1615/Int.J.UncertaintyQuantification.2021034623} \end{APACrefDOI}
\PrintBackRefs{\CurrentBib}

\bibitem [\protect \citeauthoryear {%
Chu%
\ \protect \BOthers {.}}{%
Chu%
\ \protect \BOthers {.}}{%
{\protect \APACyear {2017}}%
}]{%
chu_neural_2017}
\APACinsertmetastar {%
chu_neural_2017}%
\begin{APACrefauthors}%
Chu, X.%
, Bortnik, J.%
, Li, W.%
, Ma, Q.%
, Denton, R.%
, Yue, C.%
\BDBL {}Menietti, J.%
\end{APACrefauthors}%
\unskip\
\newblock
\APACrefYearMonthDay{2017}{{\APACmonth{09}}}{}.
\newblock
{\BBOQ}\APACrefatitle {A neural network model of three-dimensional dynamic electron density in the inner magnetosphere} {A neural network model of three-dimensional dynamic electron density in the inner magnetosphere}.{\BBCQ}
\newblock
\APACjournalVolNumPages{Journal of Geophysical Research: Space Physics}{122}{9}{9183--9197}.
\newblock
\begin{APACrefURL} [{2025-05-07}]\url{https://doi.org/10.1002/2017JA024464} \end{APACrefURL}
\newblock
\begin{APACrefDOI} \doi{10.1002/2017JA024464} \end{APACrefDOI}
\PrintBackRefs{\CurrentBib}

\bibitem [\protect \citeauthoryear {%
Chu%
, Jia%
, McPherron%
, Li%
\BCBL {}\ \BBA {} Bortnik%
}{%
Chu%
\ \protect \BOthers {.}}{%
{\protect \APACyear {2025}}%
}]{%
chu_imbalanced_2025}
\APACinsertmetastar {%
chu_imbalanced_2025}%
\begin{APACrefauthors}%
Chu, X.%
, Jia, L.%
, McPherron, R\BPBI L.%
, Li, X.%
\BCBL {}\ \BBA {} Bortnik, J.%
\end{APACrefauthors}%
\unskip\
\newblock
\APACrefYearMonthDay{2025}{}{}.
\newblock
{\BBOQ}\APACrefatitle {Imbalanced {Regression} {Artificial} {Neural} {Network} {Model} for {Auroral} {Electrojet} {Indices} ({IRANNA}): {Can} {We} {Predict} {Strong} {Events}?} {Imbalanced {Regression} {Artificial} {Neural} {Network} {Model} for {Auroral} {Electrojet} {Indices} ({IRANNA}): {Can} {We} {Predict} {Strong} {Events}?}{\BBCQ}
\newblock
\APACjournalVolNumPages{Space Weather}{23}{5}{e2024SW004236}.
\newblock
\begin{APACrefURL} [{2025-08-14}]\url{https://onlinelibrary.wiley.com/doi/abs/10.1029/2024SW004236} \end{APACrefURL}
\newblock
\begin{APACrefDOI} \doi{10.1029/2024SW004236} \end{APACrefDOI}
\PrintBackRefs{\CurrentBib}

\bibitem [\protect \citeauthoryear {%
Cornwall%
, Coroniti%
\BCBL {}\ \BBA {} Thorne%
}{%
Cornwall%
\ \protect \BOthers {.}}{%
{\protect \APACyear {1970}}%
}]{%
cornwall_turbulent_1970}
\APACinsertmetastar {%
cornwall_turbulent_1970}%
\begin{APACrefauthors}%
Cornwall, J\BPBI M.%
, Coroniti, F\BPBI V.%
\BCBL {}\ \BBA {} Thorne, R\BPBI M.%
\end{APACrefauthors}%
\unskip\
\newblock
\APACrefYearMonthDay{1970}{}{}.
\newblock
{\BBOQ}\APACrefatitle {Turbulent loss of ring current protons} {Turbulent loss of ring current protons}.{\BBCQ}
\newblock
\APACjournalVolNumPages{Journal of Geophysical Research (1896-1977)}{75}{25}{4699--4709}.
\newblock
\begin{APACrefURL} [{2026-03-12}]\url{https://onlinelibrary.wiley.com/doi/abs/10.1029/JA075i025p04699} \end{APACrefURL}
\newblock
\begin{APACrefDOI} \doi{10.1029/JA075i025p04699} \end{APACrefDOI}
\PrintBackRefs{\CurrentBib}

\bibitem [\protect \citeauthoryear {%
Evans%
}{%
Evans%
}{%
{\protect \APACyear {1973}}%
}]{%
evans_seasonal_1973}
\APACinsertmetastar {%
evans_seasonal_1973}%
\begin{APACrefauthors}%
Evans, J\BPBI V.%
\end{APACrefauthors}%
\unskip\
\newblock
\APACrefYearMonthDay{1973}{}{}.
\newblock
{\BBOQ}\APACrefatitle {Seasonal and sunspot cycle variations of {F} {Region} electron temperatures and protonospheric heat fluxes} {Seasonal and sunspot cycle variations of {F} {Region} electron temperatures and protonospheric heat fluxes}.{\BBCQ}
\newblock
\APACjournalVolNumPages{Journal of Geophysical Research (1896-1977)}{78}{13}{2344--2349}.
\newblock
\begin{APACrefURL} [{2025-08-13}]\url{https://onlinelibrary.wiley.com/doi/abs/10.1029/JA078i013p02344} \end{APACrefURL}
\newblock
\begin{APACrefDOI} \doi{10.1029/JA078i013p02344} \end{APACrefDOI}
\PrintBackRefs{\CurrentBib}

\bibitem [\protect \citeauthoryear {%
Ferradas%
\ \protect \BOthers {.}}{%
Ferradas%
\ \protect \BOthers {.}}{%
{\protect \APACyear {2023}}%
}]{%
ferradas_effects_2023}
\APACinsertmetastar {%
ferradas_effects_2023}%
\begin{APACrefauthors}%
Ferradas, C\BPBI P.%
, Fok, M\BHBI C.%
, Maruyama, N.%
, Henderson, M\BPBI G.%
, Califf, S.%
, Thaller, S\BPBI A.%
\BCBL {}\ \BBA {} Kress, B\BPBI T.%
\end{APACrefauthors}%
\unskip\
\newblock
\APACrefYearMonthDay{2023}{{\APACmonth{11}}}{}.
\newblock
{\BBOQ}\APACrefatitle {The effects of particle injections on the ring current development during the 7-8 {September} 2017 geomagnetic storm} {The effects of particle injections on the ring current development during the 7-8 {September} 2017 geomagnetic storm}.{\BBCQ}
\newblock
\APACjournalVolNumPages{Frontiers in Astronomy and Space Sciences}{10}{}{}.
\newblock
\begin{APACrefURL} [{2025-08-13}]\url{https://www.frontiersin.org/journals/astronomy-and-space-sciences/articles/10.3389/fspas.2023.1278820/full} \end{APACrefURL}
\newblock
\begin{APACrefDOI} \doi{10.3389/fspas.2023.1278820} \end{APACrefDOI}
\PrintBackRefs{\CurrentBib}

\bibitem [\protect \citeauthoryear {%
Goldstein%
}{%
Goldstein%
}{%
{\protect \APACyear {2006}}%
}]{%
goldstein_plasmasphere_2006}
\APACinsertmetastar {%
goldstein_plasmasphere_2006}%
\begin{APACrefauthors}%
Goldstein, J.%
\end{APACrefauthors}%
\unskip\
\newblock
\APACrefYearMonthDay{2006}{{\APACmonth{06}}}{}.
\newblock
{\BBOQ}\APACrefatitle {Plasmasphere {Response}: {Tutorial} and {Review} of {Recent} {Imaging} {Results}} {Plasmasphere {Response}: {Tutorial} and {Review} of {Recent} {Imaging} {Results}}.{\BBCQ}
\newblock
\APACjournalVolNumPages{Space Science Reviews}{124}{1}{203--216}.
\newblock
\begin{APACrefURL} [{2025-08-13}]\url{https://doi.org/10.1007/s11214-006-9105-y} \end{APACrefURL}
\newblock
\begin{APACrefDOI} \doi{10.1007/s11214-006-9105-y} \end{APACrefDOI}
\PrintBackRefs{\CurrentBib}

\bibitem [\protect \citeauthoryear {%
Hasegawa%
\ \BBA {} Mima%
}{%
Hasegawa%
\ \BBA {} Mima%
}{%
{\protect \APACyear {1978}}%
}]{%
hasegawa_anomalous_1978}
\APACinsertmetastar {%
hasegawa_anomalous_1978}%
\begin{APACrefauthors}%
Hasegawa, A.%
\BCBT {}\ \BBA {} Mima, K.%
\end{APACrefauthors}%
\unskip\
\newblock
\APACrefYearMonthDay{1978}{}{}.
\newblock
{\BBOQ}\APACrefatitle {Anomalous transport produced by kinetic {Alfvén} wave turbulence} {Anomalous transport produced by kinetic {Alfvén} wave turbulence}.{\BBCQ}
\newblock
\APACjournalVolNumPages{Journal of Geophysical Research: Space Physics}{83}{A3}{1117--1123}.
\newblock
\begin{APACrefURL} [{2026-03-12}]\url{https://onlinelibrary.wiley.com/doi/abs/10.1029/JA083iA03p01117} \end{APACrefURL}
\newblock
\begin{APACrefDOI} \doi{10.1029/JA083iA03p01117} \end{APACrefDOI}
\PrintBackRefs{\CurrentBib}

\bibitem [\protect \citeauthoryear {%
Hu%
\ \protect \BOthers {.}}{%
Hu%
\ \protect \BOthers {.}}{%
{\protect \APACyear {2020}}%
}]{%
hu_deep_2020}
\APACinsertmetastar {%
hu_deep_2020}%
\begin{APACrefauthors}%
Hu, A.%
, Carter, B.%
, Currie, J.%
, Norman, R.%
, Wu, S.%
\BCBL {}\ \BBA {} Zhang, K.%
\end{APACrefauthors}%
\unskip\
\newblock
\APACrefYearMonthDay{2020}{{\APACmonth{02}}}{}.
\newblock
{\BBOQ}\APACrefatitle {A {Deep} {Neural} {Network} {Model} of {Global} {Topside} {Electron} {Temperature} {Using} {Incoherent} {Scatter} {Radars} and {Its} {Application} to {GNSS} {Radio} {Occultation}} {A {Deep} {Neural} {Network} {Model} of {Global} {Topside} {Electron} {Temperature} {Using} {Incoherent} {Scatter} {Radars} and {Its} {Application} to {GNSS} {Radio} {Occultation}}.{\BBCQ}
\newblock
\APACjournalVolNumPages{Journal of Geophysical Research: Space Physics}{125}{2}{e2019JA027263}.
\newblock
\begin{APACrefURL} [{2025-05-15}]\url{https://doi.org/10.1029/2019JA027263} \end{APACrefURL}
\newblock
\begin{APACrefDOI} \doi{10.1029/2019JA027263} \end{APACrefDOI}
\PrintBackRefs{\CurrentBib}

\bibitem [\protect \citeauthoryear {%
Kaplan%
\ \protect \BOthers {.}}{%
Kaplan%
\ \protect \BOthers {.}}{%
{\protect \APACyear {2020}}%
}]{%
kaplan_scaling_2020}
\APACinsertmetastar {%
kaplan_scaling_2020}%
\begin{APACrefauthors}%
Kaplan, J.%
, McCandlish, S.%
, Henighan, T.%
, Brown, T\BPBI B.%
, Chess, B.%
, Child, R.%
\BDBL {}Amodei, D.%
\end{APACrefauthors}%
\unskip\
\newblock
\APACrefYearMonthDay{2020}{{\APACmonth{01}}}{}.
\newblock
\APACrefbtitle {Scaling {Laws} for {Neural} {Language} {Models}.} {Scaling {Laws} for {Neural} {Language} {Models}.}
\newblock
\APACaddressPublisher{}{arXiv}.
\newblock
\begin{APACrefURL} [{2025-08-14}]\url{http://arxiv.org/abs/2001.08361} \end{APACrefURL}
\newblock
\APACrefnote{arXiv:2001.08361 [cs]}
\newblock
\begin{APACrefDOI} \doi{10.48550/arXiv.2001.08361} \end{APACrefDOI}
\PrintBackRefs{\CurrentBib}

\bibitem [\protect \citeauthoryear {%
Khazanov%
, Kang%
\BCBL {}\ \BBA {} Glocer%
}{%
Khazanov%
\ \protect \BOthers {.}}{%
{\protect \APACyear {2024}}%
}]{%
khazanov_inertia_2024}
\APACinsertmetastar {%
khazanov_inertia_2024}%
\begin{APACrefauthors}%
Khazanov, G\BPBI V.%
, Kang, S\BHBI B.%
\BCBL {}\ \BBA {} Glocer, A.%
\end{APACrefauthors}%
\unskip\
\newblock
\APACrefYearMonthDay{2024}{}{}.
\newblock
{\BBOQ}\APACrefatitle {Inertia of {Ionospheric} {Conductance} {During} {Electron} {Precipitation} {Events}} {Inertia of {Ionospheric} {Conductance} {During} {Electron} {Precipitation} {Events}}.{\BBCQ}
\newblock
\APACjournalVolNumPages{Geophysical Research Letters}{51}{19}{e2024GL110813}.
\newblock
\begin{APACrefURL} [{2025-08-13}]\url{https://onlinelibrary.wiley.com/doi/abs/10.1029/2024GL110813} \end{APACrefURL}
\newblock
\begin{APACrefDOI} \doi{10.1029/2024GL110813} \end{APACrefDOI}
\PrintBackRefs{\CurrentBib}

\bibitem [\protect \citeauthoryear {%
Kitamura%
\ \protect \BOthers {.}}{%
Kitamura%
\ \protect \BOthers {.}}{%
{\protect \APACyear {2011}}%
}]{%
kitamura_solar_2011}
\APACinsertmetastar {%
kitamura_solar_2011}%
\begin{APACrefauthors}%
Kitamura, N.%
, Ogawa, Y.%
, Nishimura, Y.%
, Terada, N.%
, Ono, T.%
, Shinbori, A.%
\BDBL {}Smilauer, J.%
\end{APACrefauthors}%
\unskip\
\newblock
\APACrefYearMonthDay{2011}{}{}.
\newblock
{\BBOQ}\APACrefatitle {Solar zenith angle dependence of plasma density and temperature in the polar cap ionosphere and low-altitude magnetosphere during geomagnetically quiet periods at solar maximum} {Solar zenith angle dependence of plasma density and temperature in the polar cap ionosphere and low-altitude magnetosphere during geomagnetically quiet periods at solar maximum}.{\BBCQ}
\newblock
\APACjournalVolNumPages{Journal of Geophysical Research: Space Physics}{116}{A8}{}.
\newblock
\begin{APACrefURL} [{2025-08-13}]\url{https://onlinelibrary.wiley.com/doi/abs/10.1029/2011JA016631} \end{APACrefURL}
\newblock
\begin{APACrefDOI} \doi{10.1029/2011JA016631} \end{APACrefDOI}
\PrintBackRefs{\CurrentBib}

\bibitem [\protect \citeauthoryear {%
Kozyra%
, Nagy%
\BCBL {}\ \BBA {} Slater%
}{%
Kozyra%
\ \protect \BOthers {.}}{%
{\protect \APACyear {1997}}%
}]{%
kozyra_high-altitude_1997}
\APACinsertmetastar {%
kozyra_high-altitude_1997}%
\begin{APACrefauthors}%
Kozyra, J\BPBI U.%
, Nagy, A\BPBI F.%
\BCBL {}\ \BBA {} Slater, D\BPBI W.%
\end{APACrefauthors}%
\unskip\
\newblock
\APACrefYearMonthDay{1997}{}{}.
\newblock
{\BBOQ}\APACrefatitle {High-altitude energy source(s) for stable auroral red arcs} {High-altitude energy source(s) for stable auroral red arcs}.{\BBCQ}
\newblock
\APACjournalVolNumPages{Reviews of Geophysics}{35}{2}{155--190}.
\newblock
\begin{APACrefURL} [{2026-03-01}]\url{https://onlinelibrary.wiley.com/doi/abs/10.1029/96RG03194} \end{APACrefURL}
\newblock
\APACrefnote{\_eprint: https://agupubs.onlinelibrary.wiley.com/doi/pdf/10.1029/96RG03194}
\newblock
\begin{APACrefDOI} \doi{10.1029/96RG03194} \end{APACrefDOI}
\PrintBackRefs{\CurrentBib}

\bibitem [\protect \citeauthoryear {%
Kutiev%
, Oyama%
\BCBL {}\ \BBA {} Abe%
}{%
Kutiev%
\ \protect \BOthers {.}}{%
{\protect \APACyear {2002}}%
}]{%
kutiev_analytical_2002}
\APACinsertmetastar {%
kutiev_analytical_2002}%
\begin{APACrefauthors}%
Kutiev, I.%
, Oyama, K.%
\BCBL {}\ \BBA {} Abe, T.%
\end{APACrefauthors}%
\unskip\
\newblock
\APACrefYearMonthDay{2002}{}{}.
\newblock
{\BBOQ}\APACrefatitle {Analytical representation of the plasmasphere electron temperature distribution based on {Akebono} data} {Analytical representation of the plasmasphere electron temperature distribution based on {Akebono} data}.{\BBCQ}
\newblock
\APACjournalVolNumPages{Journal of Geophysical Research: Space Physics}{107}{A12}{SMP 24--1--SMP 24--11}.
\newblock
\begin{APACrefURL} [{2026-03-11}]\url{https://onlinelibrary.wiley.com/doi/abs/10.1029/2002JA009494} \end{APACrefURL}
\newblock
\begin{APACrefDOI} \doi{10.1029/2002JA009494} \end{APACrefDOI}
\PrintBackRefs{\CurrentBib}

\bibitem [\protect \citeauthoryear {%
Li%
\ \protect \BOthers {.}}{%
Li%
\ \protect \BOthers {.}}{%
{\protect \APACyear {2021}}%
}]{%
li_application_2021}
\APACinsertmetastar {%
li_application_2021}%
\begin{APACrefauthors}%
Li, W.%
, Zhao, D.%
, He, C.%
, Shen, Y.%
, Hu, A.%
\BCBL {}\ \BBA {} Zhang, K.%
\end{APACrefauthors}%
\unskip\
\newblock
\APACrefYearMonthDay{2021}{{\APACmonth{03}}}{}.
\newblock
{\BBOQ}\APACrefatitle {Application of a {Multi}-{Layer} {Artificial} {Neural} {Network} in a 3-{D} {Global} {Electron} {Density} {Model} {Using} the {Long}-{Term} {Observations} of {COSMIC}, {Fengyun}-{3C}, and {Digisonde}} {Application of a {Multi}-{Layer} {Artificial} {Neural} {Network} in a 3-{D} {Global} {Electron} {Density} {Model} {Using} the {Long}-{Term} {Observations} of {COSMIC}, {Fengyun}-{3C}, and {Digisonde}}.{\BBCQ}
\newblock
\APACjournalVolNumPages{Space Weather}{19}{3}{e2020SW002605}.
\newblock
\begin{APACrefURL} [{2025-05-15}]\url{https://doi.org/10.1029/2020SW002605} \end{APACrefURL}
\newblock
\begin{APACrefDOI} \doi{10.1029/2020SW002605} \end{APACrefDOI}
\PrintBackRefs{\CurrentBib}

\bibitem [\protect \citeauthoryear {%
Liemohn%
\ \protect \BOthers {.}}{%
Liemohn%
\ \protect \BOthers {.}}{%
{\protect \APACyear {2000}}%
}]{%
liemohn_ring_2000}
\APACinsertmetastar {%
liemohn_ring_2000}%
\begin{APACrefauthors}%
Liemohn, M\BPBI W.%
, Kozyra, J\BPBI U.%
, Richards, P\BPBI G.%
, Khazanov, G\BPBI V.%
, Buonsanto, M\BPBI J.%
\BCBL {}\ \BBA {} Jordanova, V\BPBI K.%
\end{APACrefauthors}%
\unskip\
\newblock
\APACrefYearMonthDay{2000}{}{}.
\newblock
{\BBOQ}\APACrefatitle {Ring current heating of the thermal electrons at solar maximum} {Ring current heating of the thermal electrons at solar maximum}.{\BBCQ}
\newblock
\APACjournalVolNumPages{Journal of Geophysical Research: Space Physics}{105}{A12}{27767--27776}.
\newblock
\begin{APACrefURL} [{2025-08-13}]\url{https://onlinelibrary.wiley.com/doi/abs/10.1029/2000JA000088} \end{APACrefURL}
\newblock
\begin{APACrefDOI} \doi{10.1029/2000JA000088} \end{APACrefDOI}
\PrintBackRefs{\CurrentBib}

\bibitem [\protect \citeauthoryear {%
Loshchilov%
\ \BBA {} Hutter%
}{%
Loshchilov%
\ \BBA {} Hutter%
}{%
{\protect \APACyear {2019}}%
}]{%
loshchilov_decoupled_2019}
\APACinsertmetastar {%
loshchilov_decoupled_2019}%
\begin{APACrefauthors}%
Loshchilov, I.%
\BCBT {}\ \BBA {} Hutter, F.%
\end{APACrefauthors}%
\unskip\
\newblock
\APACrefYearMonthDay{2019}{{\APACmonth{01}}}{}.
\newblock
\APACrefbtitle {Decoupled {Weight} {Decay} {Regularization}.} {Decoupled {Weight} {Decay} {Regularization}.}
\newblock
\APACaddressPublisher{}{arXiv}.
\newblock
\begin{APACrefURL} [{2025-05-22}]\url{http://arxiv.org/abs/1711.05101} \end{APACrefURL}
\newblock
\APACrefnote{arXiv:1711.05101 [cs]}
\newblock
\begin{APACrefDOI} \doi{10.48550/arXiv.1711.05101} \end{APACrefDOI}
\PrintBackRefs{\CurrentBib}

\bibitem [\protect \citeauthoryear {%
Mahajan%
\ \BBA {} Pandey%
}{%
Mahajan%
\ \BBA {} Pandey%
}{%
{\protect \APACyear {1979}}%
}]{%
mahajan_solar_1979}
\APACinsertmetastar {%
mahajan_solar_1979}%
\begin{APACrefauthors}%
Mahajan, K.%
\BCBT {}\ \BBA {} Pandey, V.%
\end{APACrefauthors}%
\unskip\
\newblock
\APACrefYearMonthDay{1979}{}{}.
\newblock
{\BBOQ}\APACrefatitle {Solar activity changes in the electron temperature at 1000-km altitude from the {Langmuir} probe measurements on {Isis} 1 and {Explorer} 22 satellites} {Solar activity changes in the electron temperature at 1000-km altitude from the {Langmuir} probe measurements on {Isis} 1 and {Explorer} 22 satellites}.{\BBCQ}
\newblock
\APACjournalVolNumPages{Journal of Geophysical Research: Space Physics}{84}{A10}{5885--5889}.
\newblock
\begin{APACrefURL} [{2025-08-13}]\url{https://onlinelibrary.wiley.com/doi/abs/10.1029/JA084iA10p05885} \end{APACrefURL}
\newblock
\begin{APACrefDOI} \doi{10.1029/JA084iA10p05885} \end{APACrefDOI}
\PrintBackRefs{\CurrentBib}

\bibitem [\protect \citeauthoryear {%
Nair%
\ \BBA {} Hinton%
}{%
Nair%
\ \BBA {} Hinton%
}{%
{\protect \APACyear {2010}}%
}]{%
nair_rectified_2010}
\APACinsertmetastar {%
nair_rectified_2010}%
\begin{APACrefauthors}%
Nair, V.%
\BCBT {}\ \BBA {} Hinton, G\BPBI E.%
\end{APACrefauthors}%
\unskip\
\newblock
\APACrefYearMonthDay{2010}{{\APACmonth{06}}}{}.
\newblock
{\BBOQ}\APACrefatitle {Rectified linear units improve restricted boltzmann machines} {Rectified linear units improve restricted boltzmann machines}.{\BBCQ}
\newblock
\BIn{} \APACrefbtitle {Proceedings of the 27th {International} {Conference} on {International} {Conference} on {Machine} {Learning}} {Proceedings of the 27th {International} {Conference} on {International} {Conference} on {Machine} {Learning}}\ (\BPGS\ 807--814).
\newblock
\APACaddressPublisher{Madison, WI, USA}{Omnipress}.
\PrintBackRefs{\CurrentBib}

\bibitem [\protect \citeauthoryear {%
Siddique%
, Mahmud%
, Keesee%
, Ngwira%
\BCBL {}\ \BBA {} Connor%
}{%
Siddique%
\ \protect \BOthers {.}}{%
{\protect \APACyear {2022}}%
}]{%
siddique_survey_2022}
\APACinsertmetastar {%
siddique_survey_2022}%
\begin{APACrefauthors}%
Siddique, T.%
, Mahmud, M\BPBI S.%
, Keesee, A\BPBI M.%
, Ngwira, C\BPBI M.%
\BCBL {}\ \BBA {} Connor, H.%
\end{APACrefauthors}%
\unskip\
\newblock
\APACrefYearMonthDay{2022}{{\APACmonth{01}}}{}.
\newblock
{\BBOQ}\APACrefatitle {A {Survey} of {Uncertainty} {Quantification} in {Machine} {Learning} for {Space} {Weather} {Prediction}} {A {Survey} of {Uncertainty} {Quantification} in {Machine} {Learning} for {Space} {Weather} {Prediction}}.{\BBCQ}
\newblock
\APACjournalVolNumPages{Geosciences}{12}{1}{27}.
\newblock
\begin{APACrefURL} [{2025-08-14}]\url{https://www.mdpi.com/2076-3263/12/1/27} \end{APACrefURL}
\newblock
\begin{APACrefDOI} \doi{10.3390/geosciences12010027} \end{APACrefDOI}
\PrintBackRefs{\CurrentBib}

\bibitem [\protect \citeauthoryear {%
Srivastava%
, Hinton%
, Krizhevsky%
, Sutskever%
\BCBL {}\ \BBA {} Salakhutdinov%
}{%
Srivastava%
\ \protect \BOthers {.}}{%
{\protect \APACyear {2014}}%
}]{%
srivastava_dropout_2014}
\APACinsertmetastar {%
srivastava_dropout_2014}%
\begin{APACrefauthors}%
Srivastava, N.%
, Hinton, G.%
, Krizhevsky, A.%
, Sutskever, I.%
\BCBL {}\ \BBA {} Salakhutdinov, R.%
\end{APACrefauthors}%
\unskip\
\newblock
\APACrefYearMonthDay{2014}{}{}.
\newblock
{\BBOQ}\APACrefatitle {Dropout: {A} {Simple} {Way} to {Prevent} {Neural} {Networks} from {Overfitting}} {Dropout: {A} {Simple} {Way} to {Prevent} {Neural} {Networks} from {Overfitting}}.{\BBCQ}
\newblock
\APACjournalVolNumPages{Journal of Machine Learning Research}{15}{56}{1929--1958}.
\newblock
\begin{APACrefURL} [{2025-05-22}]\url{http://jmlr.org/papers/v15/srivastava14a.html} \end{APACrefURL}
\PrintBackRefs{\CurrentBib}

\bibitem [\protect \citeauthoryear {%
Tang%
\ \BBA {} Agrawal%
}{%
Tang%
\ \BBA {} Agrawal%
}{%
{\protect \APACyear {2020}}%
}]{%
tang_discretizing_2020}
\APACinsertmetastar {%
tang_discretizing_2020}%
\begin{APACrefauthors}%
Tang, Y.%
\BCBT {}\ \BBA {} Agrawal, S.%
\end{APACrefauthors}%
\unskip\
\newblock
\APACrefYearMonthDay{2020}{{\APACmonth{03}}}{}.
\newblock
\APACrefbtitle {Discretizing {Continuous} {Action} {Space} for {On}-{Policy} {Optimization}.} {Discretizing {Continuous} {Action} {Space} for {On}-{Policy} {Optimization}.}
\newblock
\APACaddressPublisher{}{arXiv}.
\newblock
\begin{APACrefURL} [{2025-05-22}]\url{http://arxiv.org/abs/1901.10500} \end{APACrefURL}
\newblock
\APACrefnote{arXiv:1901.10500 [cs]}
\newblock
\begin{APACrefDOI} \doi{10.48550/arXiv.1901.10500} \end{APACrefDOI}
\PrintBackRefs{\CurrentBib}

\bibitem [\protect \citeauthoryear {%
Titheridge%
}{%
Titheridge%
}{%
{\protect \APACyear {1998}}%
}]{%
titheridge_temperatures_1998}
\APACinsertmetastar {%
titheridge_temperatures_1998}%
\begin{APACrefauthors}%
Titheridge, J\BPBI E.%
\end{APACrefauthors}%
\unskip\
\newblock
\APACrefYearMonthDay{1998}{}{}.
\newblock
{\BBOQ}\APACrefatitle {Temperatures in the upper ionosphere and plasmasphere} {Temperatures in the upper ionosphere and plasmasphere}.{\BBCQ}
\newblock
\APACjournalVolNumPages{Journal of Geophysical Research: Space Physics}{103}{A2}{2261--2277}.
\newblock
\begin{APACrefURL} [{2025-05-22}]\url{https://onlinelibrary.wiley.com/doi/abs/10.1029/97JA03031} \end{APACrefURL}
\newblock
\begin{APACrefDOI} \doi{10.1029/97JA03031} \end{APACrefDOI}
\PrintBackRefs{\CurrentBib}

\bibitem [\protect \citeauthoryear {%
Truhlik%
, Bilitza%
\BCBL {}\ \BBA {} Triskova%
}{%
Truhlik%
\ \protect \BOthers {.}}{%
{\protect \APACyear {2012}}%
}]{%
truhlik_new_2012}
\APACinsertmetastar {%
truhlik_new_2012}%
\begin{APACrefauthors}%
Truhlik, V.%
, Bilitza, D.%
\BCBL {}\ \BBA {} Triskova, L.%
\end{APACrefauthors}%
\unskip\
\newblock
\APACrefYearMonthDay{2012}{{\APACmonth{06}}}{}.
\newblock
{\BBOQ}\APACrefatitle {A new global empirical model of the electron temperature with the inclusion of the solar activity variations for {IRI}} {A new global empirical model of the electron temperature with the inclusion of the solar activity variations for {IRI}}.{\BBCQ}
\newblock
\APACjournalVolNumPages{Earth, Planets and Space}{64}{6}{531--543}.
\newblock
\begin{APACrefURL} [{2025-08-13}]\url{https://doi.org/10.5047/eps.2011.10.016} \end{APACrefURL}
\newblock
\begin{APACrefDOI} \doi{10.5047/eps.2011.10.016} \end{APACrefDOI}
\PrintBackRefs{\CurrentBib}

\bibitem [\protect \citeauthoryear {%
Tsuruda%
\ \BBA {} Oya%
}{%
Tsuruda%
\ \BBA {} Oya%
}{%
{\protect \APACyear {1993}}%
}]{%
tsuruda_introduction_1993}
\APACinsertmetastar {%
tsuruda_introduction_1993}%
\begin{APACrefauthors}%
Tsuruda, K.%
\BCBT {}\ \BBA {} Oya, H.%
\end{APACrefauthors}%
\unskip\
\newblock
\APACrefYearMonthDay{1993}{}{}.
\newblock
{\BBOQ}\APACrefatitle {Introduction to the {Akebono} ({EXOS} {D}) {Project}} {Introduction to the {Akebono} ({EXOS} {D}) {Project}}.{\BBCQ}
\newblock
\APACjournalVolNumPages{Journal of Geophysical Research: Space Physics}{98}{A7}{11123--11125}.
\newblock
\begin{APACrefURL} [{2026-03-11}]\url{https://onlinelibrary.wiley.com/doi/abs/10.1029/92JA02231} \end{APACrefURL}
\newblock
\begin{APACrefDOI} \doi{10.1029/92JA02231} \end{APACrefDOI}
\PrintBackRefs{\CurrentBib}

\bibitem [\protect \citeauthoryear {%
Usanova%
, Delzanno%
\BCBL {}\ \BBA {} Maruyama%
}{%
Usanova%
\ \protect \BOthers {.}}{%
{\protect \APACyear {2025}}%
}]{%
usanova_role_2025}
\APACinsertmetastar {%
usanova_role_2025}%
\begin{APACrefauthors}%
Usanova, M\BPBI E.%
, Delzanno, G\BPBI L.%
\BCBL {}\ \BBA {} Maruyama, N.%
\end{APACrefauthors}%
\unskip\
\newblock
\APACrefYearMonthDay{2025}{{\APACmonth{05}}}{}.
\newblock
{\BBOQ}\APACrefatitle {The role of wave-particle interactions in cold and warm plasma heating} {The role of wave-particle interactions in cold and warm plasma heating}.{\BBCQ}
\newblock
\APACjournalVolNumPages{Frontiers in Astronomy and Space Sciences}{12}{}{}.
\newblock
\begin{APACrefURL} [{2025-08-13}]\url{https://www.frontiersin.org/journals/astronomy-and-space-sciences/articles/10.3389/fspas.2025.1573386/full} \end{APACrefURL}
\newblock
\begin{APACrefDOI} \doi{10.3389/fspas.2025.1573386} \end{APACrefDOI}
\PrintBackRefs{\CurrentBib}

\bibitem [\protect \citeauthoryear {%
Webb%
\ \BBA {} Essex%
}{%
Webb%
\ \BBA {} Essex%
}{%
{\protect \APACyear {2003}}%
}]{%
webb_modifications_2003}
\APACinsertmetastar {%
webb_modifications_2003}%
\begin{APACrefauthors}%
Webb, P\BPBI A.%
\BCBT {}\ \BBA {} Essex, E\BPBI A.%
\end{APACrefauthors}%
\unskip\
\newblock
\APACrefYearMonthDay{2003}{}{}.
\newblock
{\BBOQ}\APACrefatitle {Modifications to the {Titheridge} upper ionosphere and plasmasphere temperature model} {Modifications to the {Titheridge} upper ionosphere and plasmasphere temperature model}.{\BBCQ}
\newblock
\APACjournalVolNumPages{Journal of Geophysical Research: Space Physics}{108}{A10}{}.
\newblock
\begin{APACrefURL} [{2025-05-22}]\url{https://onlinelibrary.wiley.com/doi/abs/10.1029/2002JA009754} \end{APACrefURL}
\newblock
\begin{APACrefDOI} \doi{10.1029/2002JA009754} \end{APACrefDOI}
\PrintBackRefs{\CurrentBib}

\bibitem [\protect \citeauthoryear {%
Wing%
\ \protect \BOthers {.}}{%
Wing%
\ \protect \BOthers {.}}{%
{\protect \APACyear {2005}}%
}]{%
wing_kp_2005}
\APACinsertmetastar {%
wing_kp_2005}%
\begin{APACrefauthors}%
Wing, S.%
, Johnson, J\BPBI R.%
, Jen, J.%
, Meng, C\BHBI I.%
, Sibeck, D\BPBI G.%
, Bechtold, K.%
\BDBL {}Takahashi, K.%
\end{APACrefauthors}%
\unskip\
\newblock
\APACrefYearMonthDay{2005}{{\APACmonth{04}}}{}.
\newblock
{\BBOQ}\APACrefatitle {Kp forecast models} {Kp forecast models}.{\BBCQ}
\newblock
\APACjournalVolNumPages{Journal of Geophysical Research: Space Physics}{110}{A4}{}.
\newblock
\begin{APACrefURL} [{2025-05-15}]\url{https://doi.org/10.1029/2004JA010500} \end{APACrefURL}
\newblock
\begin{APACrefDOI} \doi{10.1029/2004JA010500} \end{APACrefDOI}
\PrintBackRefs{\CurrentBib}

\bibitem [\protect \citeauthoryear {%
Zhan%
, Doostan%
, Sutton%
\BCBL {}\ \BBA {} Fang%
}{%
Zhan%
\ \protect \BOthers {.}}{%
{\protect \APACyear {2024}}%
}]{%
zhan_quantifying_2024}
\APACinsertmetastar {%
zhan_quantifying_2024}%
\begin{APACrefauthors}%
Zhan, W.%
, Doostan, A.%
, Sutton, E.%
\BCBL {}\ \BBA {} Fang, T\BHBI W.%
\end{APACrefauthors}%
\unskip\
\newblock
\APACrefYearMonthDay{2024}{}{}.
\newblock
{\BBOQ}\APACrefatitle {Quantifying {Uncertainties} in the {Quiet}-{Time} {Ionosphere}-{Thermosphere} {Using} {WAM}-{IPE}} {Quantifying {Uncertainties} in the {Quiet}-{Time} {Ionosphere}-{Thermosphere} {Using} {WAM}-{IPE}}.{\BBCQ}
\newblock
\APACjournalVolNumPages{Space Weather}{22}{2}{e2023SW003665}.
\newblock
\begin{APACrefURL} [{2025-08-14}]\url{https://onlinelibrary.wiley.com/doi/abs/10.1029/2023SW003665} \end{APACrefURL}
\newblock
\begin{APACrefDOI} \doi{10.1029/2023SW003665} \end{APACrefDOI}
\PrintBackRefs{\CurrentBib}

\bibitem [\protect \citeauthoryear {%
Zhelavskaya%
, Shprits%
\BCBL {}\ \BBA {} Spasojević%
}{%
Zhelavskaya%
\ \protect \BOthers {.}}{%
{\protect \APACyear {2017}}%
}]{%
zhelavskaya_empirical_2017}
\APACinsertmetastar {%
zhelavskaya_empirical_2017}%
\begin{APACrefauthors}%
Zhelavskaya, I\BPBI S.%
, Shprits, Y\BPBI Y.%
\BCBL {}\ \BBA {} Spasojević, M.%
\end{APACrefauthors}%
\unskip\
\newblock
\APACrefYearMonthDay{2017}{{\APACmonth{11}}}{}.
\newblock
{\BBOQ}\APACrefatitle {Empirical {Modeling} of the {Plasmasphere} {Dynamics} {Using} {Neural} {Networks}} {Empirical {Modeling} of the {Plasmasphere} {Dynamics} {Using} {Neural} {Networks}}.{\BBCQ}
\newblock
\APACjournalVolNumPages{Journal of Geophysical Research: Space Physics}{122}{11}{11,227--11,244}.
\newblock
\begin{APACrefURL} [{2025-05-15}]\url{https://doi.org/10.1002/2017JA024406} \end{APACrefURL}
\newblock
\begin{APACrefDOI} \doi{10.1002/2017JA024406} \end{APACrefDOI}
\PrintBackRefs{\CurrentBib}

\end{thebibliography}

\end{document}